\documentclass[a4paper,fleqn]{cas-sc}

\usepackage[numbers]{natbib}
\usepackage{multicol}
\usepackage{caption}
\usepackage{subcaption}
\usepackage{setspace}
\usepackage{textcomp}
\usepackage{mathtools}
\usepackage{amsmath}
\usepackage{amssymb}
\usepackage{xurl}
\usepackage{rotating}
\usepackage{lscape}
\usepackage{stfloats}
\usepackage{mathrsfs}
\usepackage{booktabs}
\usepackage{hyperref}

\begin{document}
\let\WriteBookmarks\relax
\def\floatpagepagefraction{1}
\def\textpagefraction{.001}

\shorttitle{Benchmarking Heritability Estimation Strategies}
\shortauthors{Muneeb and Ascher}

\title[mode=title]{Benchmarking Heritability Estimation Strategies Across 86 Configurations and Their Downstream Effect on Polygenic Risk Score Performance}

\author[1,2]{Muhammad Muneeb}[auid=001,
                               bioid=1,
                               orcid=0000-0002-6506-4430]
\fnmark[1]
\ead{m.muneeb@uq.edu.au}
\ead[url]{https://orcid.org/0000-0002-6506-4430}
\credit{Conceptualization, Methodology, Software, Formal analysis, Data curation, Writing -- Original draft preparation}

\affiliation[1]{organization={School of Chemistry and Molecular Biosciences, The University of Queensland},
                addressline={},
                city={Brisbane},
                postcode={4067},
                state={Queensland},
                country={Australia}}

\author[1,2]{David B. Ascher}[auid=002,
                               bioid=2,
                               orcid=0000-0003-2948-2413]
\cormark[1]
\fnmark[2]
\ead{d.ascher@uq.edu.au}
\ead[url]{https://orcid.org/0000-0003-2948-2413}
\credit{Conceptualization, Methodology, Supervision, Writing -- Review and editing}

\affiliation[2]{organization={Computational Biology and Clinical Informatics, Baker Heart and Diabetes Institute},
                addressline={Commercial Road},
                city={Melbourne},
                postcode={3004},
                state={Victoria},
                country={Australia}}

\cortext[cor1]{Corresponding author: David B. Ascher, d.ascher@uq.edu.au}

\fntext[fn1]{Muhammad Muneeb performed all computational analyses and drafted the manuscript.}
\fntext[fn2]{David B. Ascher supervised the project and contributed to study design and manuscript revision.}

\nonumnote{The authors declare no competing interests. UK Biobank data were accessed under application ID 50000. All analyses were conducted on participants of European ancestry. The authors confirm that ethics approval was not required for this secondary analysis of existing data.}

\begin{abstract}
\textbf{Objective:} SNP heritability estimates vary substantially across estimation strategies, yet the downstream consequences for polygenic risk score (PRS) construction remain poorly characterised. We systematically benchmarked heritability estimation configurations and assessed their propagation into downstream PRS performance. 
\textbf{Methods:} We benchmarked 86 heritability-estimation configurations spanning six tool families (GEMMA, GCTA, LDAK, DPR, LDSC, SumHer) and ten method groups across 10 UK Biobank phenotypes, yielding 844 configuration-level estimates. Each estimate was propagated into GCTA-SBLUP and LDpred2-lassosum2 PRS frameworks and evaluated across five cross-validation folds using null, PRS-only, and full models. Eleven binary analytical contrasts were tested using Mann--Whitney U tests to identify drivers of heritability variability. 
\textbf{Results:} Heritability ranged from $-0.862$ to $2.735$ (mean $= 0.134$, SD $= 0.284$), with 133 of 844 estimates (15.8\%) negative and concentrated in unconstrained estimation regimes. Ten of eleven analytical contrasts significantly affected heritability magnitude, with algorithm choice and GRM standardisation showing the largest effects. Despite this upstream variability, downstream PRS test performance was only weakly coupled to heritability magnitude: pooled Pearson correlations between $h^2$ and test AUC were $r = -0.023$ for GCTA-SBLUP and $r = +0.014$ for LDpred2-lassosum2 (both non-significant). 
\textbf{Conclusion:} SNP heritability is best interpreted as a configuration-sensitive modelling parameter rather than a universally stable scalar input. Heritability estimates should always be reported alongside their full estimation specification, and downstream PRS performance is comparatively robust to moderate variation in the heritability input.
\end{abstract}

\begin{graphicalabstract}
\centering
\includegraphics[width=13cm,height=5cm,keepaspectratio]{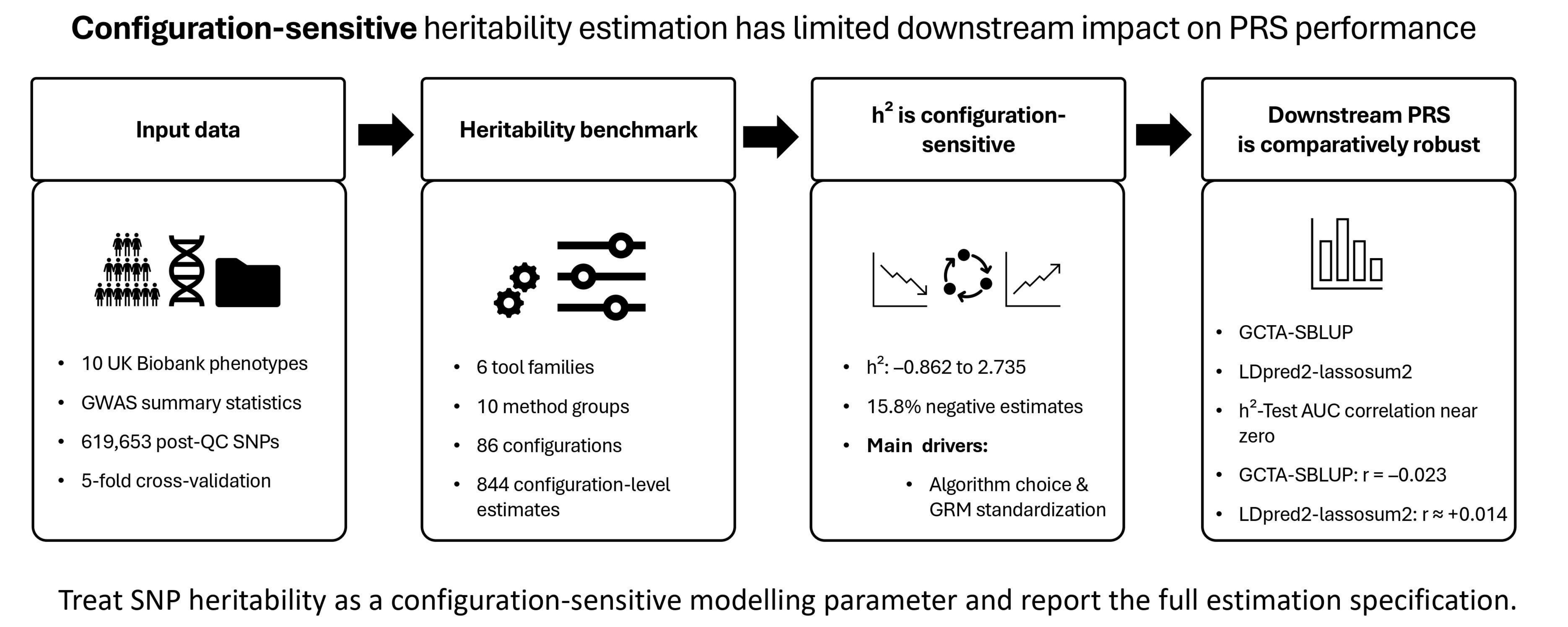}
\end{graphicalabstract}


\maketitle

\section{Introduction}
Heritability ($h^2$) measures the proportion of phenotypic variance attributable to additive genetic effects within a population \cite{Mayhew2017}. Heritability estimates derived from genotype data or Genome--Wide Association Study (GWAS) summary statistics are now routinely used as inputs to downstream statistical genetic workflows, most importantly polygenic risk score (PRS) construction. In practice, however, heritability is not a single fixed quantity returned identically across analysis pipelines.

In practice, heritability estimates are influenced by multiple factors, including gene--environment interaction, population structure, sample size, and the genetic architecture of the trait \cite{Zuk2012,Browning2011,Mayhew2017}. More importantly for applied analyses, the estimated value depends strongly on the estimation strategy, input data type, quality-control procedure, and construction of the genetic relatedness matrix (GRM) \cite{ZHU20201557,2024}. As a result, different estimation regimes applied to the same data can produce substantially different $h^2$ values. In the genomics era, heritability is commonly estimated from genotype data or GWAS summary statistics \cite{Manolio2009,Lee2011,Li2023,Srivastava2023}. Using genotype data, heritability is typically estimated through a linear mixed model:
\begin{equation}
\mathbf{y} = \mathbf{X}\beta + \epsilon
\label{equation4}
\end{equation}
where $\mathbf{y}$ is the $m \times 1$ vector of phenotypes, $\mathbf{X}$ is the $m \times n$ genotype matrix, $\beta$ is the $n \times 1$ vector of SNP effects, and $\boldsymbol{\epsilon} \sim \mathcal{N}(0, \mathbf{I}\sigma_e^2)$. Variance components are estimated using restricted maximum likelihood (REML) \cite{Corbeil1976}, method of moments \cite{Lindsay2005}, or maximum likelihood \cite{MARDIA1984}, yielding heritability as:
\begin{equation}
h^2 = \frac{\sigma_g^2}{\sigma_g^2 + \sigma_e^2}
\label{equationh2main}
\end{equation}
where $\sigma_g^2$ is the total additive genetic variance and $\sigma_e^2$ is the residual variance. Modelling approaches span linear mixed models \cite{Heckerman2016}, generalised linear mixed models \cite{Sun2019-lc}, Bayesian approaches \cite{Nustad2018-uy}, Bayesian variable selection regression \cite{https://doi.org/10.48550/arxiv.1110.6019}, and Bayesian sparse linear mixed models \cite{Zhou2013}. Because these modelling choices jointly determine the returned $h^2$ value, different tools applied to the same data can produce substantially different estimates.

One of the most important downstream applications of heritability is constructing PRS, which aggregate genome-wide SNP effects to estimate an individual's genetic predisposition to a trait or disease \cite{Choi2020,Lewis2020,Hujoel2022,Muneeb2022,Muneeb20221}. PRS have been applied to predict conditions including type 2 diabetes, attention-deficit hyperactivity disorder, and cardiovascular disease \cite{Collister2022}. Two widely used PRS frameworks, GCTA-SBLUP and LDpred2-lassosum2, accept heritability as either an estimated or user-provided parameter. In GCTA-SBLUP, $h^2$ directly determines the shrinkage parameter $\lambda = m(1/h^2 - 1)$; in LDpred2-lassosum2, heritability is used to parameterise the regularisation grid. Differences in $h^2$ can therefore propagate directly into PRS construction and affect downstream predictive performance. Despite the widespread use of SNP heritability in statistical genetics and PRS pipelines, there is limited practical guidance on how configuration-dependent variation in $h^2$ affects downstream predictive modelling in realistic analysis settings. Most studies focus either on the statistical properties of individual estimators or on PRS performance in isolation, rather than on how variation in heritability propagates into downstream prediction.

In this study we addressed this gap by benchmarking 86 heritability-estimation configurations spanning six tool families --- GEMMA \cite{zhou2012genome}, GCTA \cite{yang2011gcta}, LDAK \cite{Speed2020}, DPR \cite{Zeng2017}, LDSC \cite{bulik2015ldsc}, and SumHer \cite{Speed2018} --- organised into ten method groups and applied to 10 UK Biobank phenotypes, yielding 844 configuration-level estimates. Each $h^2$ estimate was propagated into GCTA-SBLUP and LDpred2-lassosum2 \cite{Priv2020} and evaluated across five cross-validation folds. We characterised which configurations produced negative heritability values and why, quantified how SNP count and preprocessing choices affected estimates, tested whether heritability magnitude predicted downstream PRS accuracy, and quantified how sensitive downstream PRS performance is to heritability configuration across the full 86-configuration space.

\begin{table}[h!]
\caption{\textbf{Statement of Significance}}
\begin{tabular}{p{3.5cm}p{8.5cm}}
\toprule
\textbf{Problem or Issue} & SNP heritability estimates vary substantially across estimation strategies, yet practical guidance on how this variation propagates into polygenic risk score construction is lacking. \\
\midrule
\textbf{What is Already Known} & Individual heritability estimators differ in their statistical assumptions and properties, but systematic benchmarking across a broad range of configurations and their downstream PRS consequences has not been conducted. \\
\midrule
\textbf{What this Paper Adds} & A benchmark of 86 estimation configurations across six tool families and 10 UK Biobank phenotypes showing that heritability is highly configuration-sensitive but downstream PRS performance is comparatively robust to this variability. \\
\midrule
\textbf{Who Would Benefit} & Statistical geneticists and clinical researchers constructing PRS pipelines who need practical guidance on heritability estimation strategy selection and reporting. \\
\bottomrule
\end{tabular}
\end{table}

\section{Related Work}
Heritability estimation from genome-wide data has been approached through several methodological traditions. Restricted maximum likelihood (REML) methods, implemented in tools such as GCTA \citep{yang2011gcta} and GEMMA \citep{zhou2012genome}, estimate variance components by fitting linear mixed models to individual-level genotype data and impose a non-negativity constraint on heritability during optimisation \citep{Mayhew2017}. Haseman--Elston (HE) regression offers a computationally lighter alternative that regresses phenotypic similarity on genetic similarity but does not enforce non-negativity and can return negative estimates under low signal-to-noise conditions \citep{ZHU20201557}. LD score regression (LDSC) \citep{bulik2015ldsc} and its extension SumHer \citep{Speed2018} operate on GWAS summary statistics rather than individual-level data, using the relationship between test statistics and linkage disequilibrium scores to estimate heritability without access to raw genotypes. LDAK \citep{Speed2020} extends this framework by incorporating SNP-specific weightings that account for differences in tagging efficiency across the genome. Bayesian approaches including DPR \citep{Zeng2017} and LDpred2 \citep{Priv2020} model the full posterior distribution of SNP effects and can propagate uncertainty into downstream applications. Despite the breadth of available estimators, systematic comparisons across a large number of configurations and their downstream consequences for PRS construction remain limited. Prior work has examined individual estimator properties \citep{ZHU20201557,2024} or compared a small number of tools on specific traits, but has not evaluated how configuration-level variability in heritability propagates into PRS performance across a standardised multi-phenotype benchmark with cross-validation-based robustness assessment. This study addresses that gap directly. 

\section{Methods}
\subsection{Study cohort, phenotype selection, and sample characteristics}
Genotype data were extracted from the UK Biobank for 14 candidate phenotypes and matched to corresponding GWAS summary statistics files obtained from the GWAS Catalog (\url{https://www.ebi.ac.uk/gwas/}). Quality controls were applied to both datasets independently, and three phenotypes with insufficient SNP overlap were removed, leaving 10 phenotypes for downstream heritability benchmarking and PRS evaluation.

GWAS files were standardised and harmonised using GWASPokerforPRS (\url{https://github.com/MuhammadMuneeb007/GWASPokerforPRS}). Genotype quality control was applied using PLINK, with variants filtered on minor allele frequency (MAF $> 0.01$), Hardy--Weinberg equilibrium ($p > 1 \times 10^{-6}$), genotype missingness ($< 0.1$), and individual missingness ($< 0.1$). Relatedness was controlled using a kinship coefficient cutoff of 0.125. GWAS summary statistics were additionally filtered to retain SNPs with MAF $> 0.01$ and imputation information score (INFO $> 0.8$); ambiguous SNPs with complementary alleles (C/G or A/T) were excluded to prevent strand-assignment errors. Three phenotypes were excluded because the number of shared variants was insufficient for reliable estimation: blood pressure medication ($n = 56$ common SNPs), cholesterol-lowering medication ($n = 56$), and hayfever/allergic rhinitis ($n = 56$). The remaining 10 phenotypes were retained for heritability benchmarking and PRS evaluation. 

The final post-QC genotype dataset comprised 619,653 SNPs and 135 covariates, including nuclear magnetic resonance (NMR) metabolomic biomarkers and comorbid conditions, derived from the UK Biobank data. All analyses were conducted on participants of European ancestry retained after quality control, consistent with the ancestry composition of the analysed UK Biobank subset and the European LD reference resources used by LDSC and LDAK. 

% GWAS summary statistics were derived from studies that included UK Biobank participants --- notably depression \cite{Howard2018}, high cholesterol \cite{Loh2018}, hypertension \cite{GuindoMartnez2021}, irritable bowel syndrome \cite{Eijsbouts2021}, migraine \cite{Jiang2021}, and osteoarthritis \cite{McDonald2022} --- potential sample overlap between the GWAS discovery cohort and the UK Biobank target sample may partially inflate heritability and PRS performance estimates for those phenotypes. 

Sample sizes, case--control counts, post-QC SNP counts, and GWAS sources for all 10 retained phenotypes are reported in Table~\ref{tab:sample_characteristics}. 

\begin{table*}[!ht]
\centering
\caption{\textbf{Sample characteristics and GWAS sources for the 10 retained phenotypes.} $N$ is the total analytic sample after quality control. Cases and controls are reported for binary phenotypes; body mass index is a continuous trait (denoted by dashes). SNPs (QC) is the number of variants in the post-QC genotype dataset. Common SNPs is the number of variants shared between the GWAS summary statistics file and the post-QC genotype data.}
\label{tab:sample_characteristics}
\resizebox{\textwidth}{!}{%
\begin{tabular}{lrrrrrrl}
\toprule
\textbf{Phenotype} & \textbf{$N$ total} & \textbf{Cases} & \textbf{Controls} & \textbf{SNPs (QC)} & \textbf{Common SNPs} & \textbf{Trait type} & \textbf{GWAS source} \\
\midrule
Asthma                    & 733 & 115 & 618 & 619{,}653 & 2{,}867   & Binary     & \cite{Sakaue2021} \\
Body Mass Index           & 728 & —   & —   & 619{,}653 & 2{,}866   & Continuous & \cite{Sakaue2021} \\
Depression                & 733 & 114 & 619 & 619{,}653 & 545{,}218 & Binary     & \cite{Howard2018} \\
Gastro-Oesophageal Reflux & 733 &  45 & 688 & 619{,}653 & 2{,}906   & Binary     & \cite{Sakaue2021} \\
High Cholesterol          & 733 &  82 & 651 & 619{,}653 & 562{,}302 & Binary     & \cite{Loh2018} \\
Hypertension              & 733 & 188 & 545 & 619{,}653 & 619{,}653 & Binary     & \cite{GuindoMartnez2021} \\
Hypothyroidism            & 733 &  70 & 663 & 619{,}653 & 2{,}908   & Binary     & \cite{Sakaue2021} \\
Irritable Bowel Syndrome  & 733 &  78 & 655 & 619{,}653 & 568{,}535 & Binary     & \cite{Eijsbouts2021} \\
Migraine                  & 733 &  53 & 680 & 619{,}653 & 548{,}955 & Binary     & \cite{Jiang2021} \\
Osteoarthritis            & 733 &  94 & 639 & 619{,}653 & 619{,}653 & Binary     & \cite{McDonald2022} \\
\bottomrule
\end{tabular}%
}
\end{table*}

To enable within-sample estimation and held-out evaluation of heritability and PRS performance, each phenotype was divided into five cross-validation folds. For binary phenotypes, stratified five-fold cross-validation was applied to preserve case--control proportions across partitions. For body mass index, standard five-fold cross-validation was used. Each fold comprised a training partition (80\% of samples) and a held-out test partition (20\%). Heritability estimation was performed exclusively on the training fold; the resulting fold-specific $h^2$ estimate was used to parameterise the PRS model for that fold. PRS effect sizes were computed from training data and evaluated on the held-out test fold without refitting, ensuring that test-fold performance reflects genuine out-of-sample generalisation. Quality control, clumping, and pruning were applied exclusively within the training partition; the test fold was never used in any preprocessing or estimation step.

\subsection{Heritability benchmark design}
We benchmarked 86 distinct heritability estimation configurations derived from six tool families --- GEMMA \cite{zhou2012genome}, GCTA \cite{yang2011gcta}, LDAK \cite{Speed2020}, DPR \cite{Zeng2017}, LDSC \cite{bulik2015ldsc}, and SumHer \cite{Speed2018} --- organised into ten method groups. Because each configuration simultaneously varies software, input data type, algorithmic choices, and preprocessing decisions, this benchmark compares estimation strategies and configurations rather than software in isolation. The full set of configurations is described in Table~\ref{Table2}.

\begin{landscape}
\begin{table}[]
\centering
\caption{\textbf{Heritability estimation configurations benchmarked in this study.} Each row defines one method group. The Index column provides the identifier assigned to each group for reference. The Tool column lists the software used for heritability estimation. The Variants column describes the data inputs, algorithmic choices, and preprocessing decisions that distinguish configurations within each group. The GWAS and Genotype data columns indicate which input data types are required. The Covariates column indicates whether covariates were included in the estimation model. The Reference Panel column specifies whether an external LD reference resource was used. The Clumping and Pruning column indicates whether variant filtering was applied before estimation. The Models column reports the number of unique heritability configurations generated by each method group, summing to 86 configurations in total. \textbf{Method 1: LDpred2} uses LDSC internally to estimate heritability. It can restrict SNPs to HapMap3 variants or use all variants common between the GWAS file and genotype data. Clumping and pruning may be applied before passing data to LDpred2, resulting in 4 configurations. \textbf{Method 2: GCTA} accepts genotype data and optionally includes covariates and principal components. Clumping and pruning may be applied before estimation, resulting in 6 configurations. \textbf{Method 3: GEMMA} uses genotype data and constructs two types of GRM (centred and standardised). It supports two estimation algorithms: Haseman--Elston (HE) regression and REML-AI. Covariates and principal components may be included, and clumping and pruning may be applied, resulting in 24 configurations. \textbf{Method 4: GEMMA with GWAS} extends Method 3 by additionally incorporating GWAS summary statistics. It uses HE regression with genotype data, covariates, and principal components, with and without clumping and pruning, resulting in 6 configurations. \textbf{Method 5: LDSC+GEMMA} uses the REML-AI algorithm within GEMMA together with an LD score reference panel derived from the genotype data via LDSC. Clumping and pruning are applied because computing LD scores for all variants is computationally prohibitive, resulting in 3 configurations. \textbf{Method 6: DPR+GEMMA} uses DPR to construct centred and standardised GRMs, which are then passed to GEMMA with either HE regression or REML-AI. Covariates and principal components may be included, and clumping and pruning may be applied, resulting in 24 configurations. \textbf{Method 7: LDSC (EUR reference)} estimates heritability using LDSC with a precomputed European reference panel and European LD weights. Four configurations are created by varying the reference and weight file combinations. \textbf{Method 8: LDAK-Calculated} uses LDAK to compute tagging files from the genotype data under four tagging models (Human, GCTA, BLD-LDAK, Alpha) and then estimates heritability from the GWAS file, resulting in 8 configurations. \textbf{Method 9: LDAK-Precomputed} uses six precomputed tagging files provided by LDAK for the UK Biobank GBR population, covering HapMap and genotyped SNP sets under different annotation models, resulting in 6 configurations. \textbf{Method 10: LDSC (genotype-derived reference)} estimates heritability using LDSC with a reference panel and LD weights derived from the genotype data rather than an external resource, resulting in 1 configuration.}
\label{Table2}
\resizebox{\columnwidth}{!}{%
\begin{tabular}{|l|l|l|l|l|l|l|l|l|}
\hline
\textbf{Index} & \textbf{Tool} & \textbf{Variants} & \textbf{GWAS} & \textbf{Genotype data} & \textbf{Covariates} & \textbf{Reference Panel} & \textbf{Clumping and Pruning} & \textbf{Models} \\ \hline
Method 1 & LDpred2 & \begin{tabular}[c]{@{}l@{}}LDpred2\_full\\ LDpred2\_hapmap\end{tabular} & Yes & Yes & No & No & Yes/No & 4 \\ \hline
Method 2 & GCTA & (Data) Genotype/Genotype+Covariate/Genotype+Covariate+PCA & No & Yes & Yes & No & Yes/No & 6 \\ \hline
Method 3 & GEMMA & \begin{tabular}[c]{@{}l@{}}(GRM) Centred/Standardised\\ (Algorithm) HE regression/REML AI\\ (Data) Genotype/+Covariate/+Covariate+PCA\end{tabular} & No & Yes & Yes & No & Yes/No & 24 \\ \hline
Method 4 & GEMMA & \begin{tabular}[c]{@{}l@{}}(Algorithm) HE regression\\ (Data) Genotype/+Covariate/+Covariate+PCA\end{tabular} & Yes & Yes & Yes & No & Yes/No & 6 \\ \hline
Method 5 & LDSC+GEMMA & \begin{tabular}[c]{@{}l@{}}(Algorithm) REML AI\\ (Data) Genotype/+Covariate/+Covariate+PCA\end{tabular} & Yes & Yes & Yes & LD Score of Genotype data & Yes & 3 \\ \hline
Method 6 & DPR+GEMMA & \begin{tabular}[c]{@{}l@{}}(GRM) Centred/Standardised from DPR\\ (Algorithm) HE regression/REML AI\\ (Data) Genotype/+Covariate/+Covariate+PCA\end{tabular} & No & Yes & Yes & No & Yes/No & 24 \\ \hline
Method 7 & LDSC & LDSC (EUR reference panel, EUR weights) & Yes & No & No & EUR reference panel, EUR weights & No & 4 \\ \hline
Method 8 & LDAK-Calculated & LDAK tagging models (Human, GCTA, BLD-LDAK, Alpha) & Yes & Yes & No & No & Yes/No & 8 \\ \hline
Method 9 & LDAK-Precomputed & \begin{tabular}[c]{@{}l@{}}Precomputed taggings:\\ bld.ldak.hapmap.gbr.tagging\\ ldak.thin.hapmap.gbr.tagging\\ bld.ldak.lite.alpha.hapmap.gbr.tagging\\ bld.ldak.genotyped.gbr.tagging\\ ldak.thin.genotyped.gbr.tagging\\ bld.ldak.lite.alpha.genotyped.gbr.tagging\end{tabular} & Yes & No & No & No & No & 6 \\ \hline
Method 10 & LDSC & LDSC (reference panel and EUR weights from genotype data) & Yes & Yes & No & LD Score of Genotype data & Yes & 1 \\ \hline
\end{tabular}%
}
\end{table}
\end{landscape}

To characterise the stability and reliability of heritability estimates across configurations, we computed for each configuration and phenotype the mean $h^2$ averaged across five cross-validation folds, the fold-derived standard error (fold-SE $= \mathrm{SD}/\sqrt{n_{\text{folds}}}$), and the corresponding 95\% confidence interval (CI $=$ mean $h^2 \pm 1.96 \times$ fold-SE). A configuration was classified as yielding a statistically reliable positive estimate if the lower bound of the fold-based CI exceeded zero. The 10 method groups were consolidated into six method families: LDpred2, GCTA, GEMMA (Methods 3--5), DPR+GEMMA, LDSC (Methods 7 and 10), and LDAK (Methods 8--9). The coefficient of variation (CV $=$ SD\,/\,$|$mean $h^2|$) was computed per family as a scale-free measure of estimation instability. 

Negative heritability estimates were tabulated per method family. To assess the effect of SNP count on heritability magnitude, Spearman rank correlations were computed between the mean number of variants used in estimation and the mean $h^2$ estimate, both overall and separately within each method family and phenotype. Full fold-level estimates, standard errors, and 95\% confidence intervals for all 86 configurations and 10 phenotypes are provided in Supplementary Table~S1 (Heritability Estimates).

\subsection{Effect of estimation hyperparameters on heritability}
To assess the effect of individual analytical choices on heritability estimates, we extracted eleven binary hyperparameter contrasts from the 86 estimation configurations: clumping and pruning applied versus skipped; HE regression versus other algorithms; REML versus other algorithms; covariate inclusion versus exclusion; PCA inclusion versus exclusion; GWAS summary statistics used versus not; genotype data used versus not; centred versus standardised GRM; and HapMap versus genotyped tagging files for LDAK. For each contrast, distributions of mean $h^2$ estimates were compared using two-sided Mann--Whitney U tests, both pooled across all method families and separately within each family. A contrast was classified as statistically significant at $p < 0.05$. Effect size was quantified as the difference in group means ($\Delta h^2$).

\subsection{Inter-method correlation and agreement}
To assess whether different heritability estimation strategies produce consistent relative rankings of phenotypes, we computed pairwise Pearson and Spearman correlations between all ten estimation methods using their mean $h^2$ profile across the ten phenotypes. Correlations were also computed at the method-family level by averaging $h^2$ values across configurations within each family. The resulting $6 \times 6$ family-level correlation matrix was visualised as a clustered heatmap with a dendrogram computed using hierarchical clustering with average linkage and Euclidean distance. Statistical significance of each pairwise correlation was assessed using a two-sided Pearson $t$-test ($p < 0.05$). To test whether methods within the same family agreed more strongly than methods from different families, same-family and different-family pairwise Pearson correlations were compared using a two-sided Mann--Whitney U test. Method-level ranking consistency was assessed using a Friedman test across all ten methods and phenotypes.

\subsection{Matched-input controlled comparison}
To separate the contribution of the statistical estimator from preprocessing decisions, we defined a matched-input subset in which each method and phenotype was represented by the configuration retaining the largest number of variants, with no clumping preferred as a tiebreaker when variant counts were equal. Under these matched-input conditions, any remaining differences in $h^2$ across methods are attributable to the statistical estimator --- its modelling assumptions, LD handling, and variance component algorithm --- rather than to preprocessing or input data type. Matched-input estimates were compared across methods using a Kruskal--Wallis test and pairwise Pearson correlations over the ten phenotypes. Matched-input means were also compared to the full benchmark means for each method, with $\Delta = $ full mean $-$ matched mean; a positive $\Delta$ indicates that the full benchmark inflates $h^2$ relative to the maximum-variant baseline.

\subsection{Polygenic risk score construction}
Two PRS frameworks were evaluated: GCTA-SBLUP and LDpred2-lassosum2. For each heritability estimation configuration, the fold-specific mean $h^2$ estimate was used to parameterise the corresponding PRS model, yielding one PRS model per heritability configuration per fold. This design allowed us to examine directly how variation in estimated heritability propagates into downstream predictive performance.

\subsubsection{GCTA-SBLUP}
GCTA-SBLUP requires a shrinkage parameter $\lambda$ computed from the heritability estimate and the number of SNPs retained after quality control:
\begin{equation}
\lambda = m \left( \frac{1}{h^2} - 1 \right)
\label{equation_lambda}
\end{equation}
where $m$ is the number of SNPs used in estimation and $h^2$ is the fold-specific heritability estimate for that configuration. GCTA uses $\lambda$ together with the GWAS summary statistics and genotype data to compute shrinkage-adjusted SNP effect sizes $\hat{\beta}$ via Single-step Best Linear Unbiased Prediction (SBLUP) \cite{Chung2021,Yang2016}:
\begin{equation}
\hat{\beta} = \left( \mathbf{G} + \mathbf{I} \lambda \right)^{-1} \mathbf{X}^T \mathbf{y}
\end{equation}
where $\mathbf{G}$ is the genomic relationship matrix, $\mathbf{I}$ is the identity matrix, $\mathbf{X}$ is the genotype matrix, and $\mathbf{y}$ is the phenotype vector.

\subsubsection{LDpred2-lassosum2}
LDpred2-lassosum2 \cite{Priv2020} uses the LD matrix computed from the training genotype data together with GWAS summary statistics to estimate posterior SNP effects across a grid of regularisation parameters ($\lambda$, $\delta$). The LD matrix was computed chromosome by chromosome using \texttt{snp\_cor} from the \texttt{bigsnpr} package with size parameter 200, $\alpha = 1$, and $r^2$ threshold 0.1. Heritability was estimated internally via LD score regression using \texttt{snp\_ldsc} applied to the fold-specific training data. The lassosum2 grid spanned four values of $\delta$ ($0.001, 0.01, 0.1, 1$) and ten values of $\lambda$, yielding 40 grid points per configuration. Two SNP sets were evaluated: HapMap3 SNPs only (\texttt{LDpred2\_hapmap}) and all SNPs common between the GWAS file and genotype data (\texttt{LDpred2\_full}).

\subsubsection{PRS scoring and model evaluation}
For both frameworks, PRS were scored using PLINK \texttt{--score}:
\begin{equation}
\mathrm{PRS} = \sum_{i=1}^{n} \hat{\beta}_i \times G_i
\label{Plinkequation}
\end{equation}
where $\hat{\beta}_i$ is the effect size for SNP $i$ and $G_i \in \{0, 1, 2\}$ is the allele dosage. Prior to scoring, genotype data were clumped and pruned (window 200 kb, $r^2 = 0.1$, $p$-value threshold 1) on the training fold only; quality control was never applied to the test fold.

\subsection{PRS evaluation design}
To assess downstream predictive performance, we fitted three models for each configuration and fold using the training data and evaluated them on the held-out test fold: (1) a null model using covariates and the first six principal components only, with no PRS, establishing the baseline predictive performance attributable to demographic and clinical factors alone; (2) a PRS-only model using the raw PRS score without covariates, isolating the raw genetic signal; and (3) a full model combining PRS with all covariates, representing the primary evaluation condition. For binary phenotypes, logistic regression was used and performance was reported as area under the receiver operating characteristic curve (AUC). For body mass index, ordinary least squares regression was used and performance was reported as explained variance ($R^2$). Models were fitted on the training fold and evaluated on the test fold without refitting. Performance metrics were averaged across the five cross-validation folds. The overfitting indicator for each configuration was computed as $\Delta = \bar{\mathrm{AUC}}_{\mathrm{train}} - \bar{\mathrm{AUC}}_{\mathrm{test}}$; configurations with lower $\Delta$ and higher test AUC were considered more generalisable.

\subsection{Exploratory comparison of configuration selection strategies}

To summarise configuration-space behaviour, we compared a delta-constrained selection heuristic with three alternative strategies: best-train selection, random selection (mean over 1000 draws), and an oracle upper bound defined by the highest observed test performance. The delta-constrained heuristic prioritised configurations with relatively small train--test gaps while favouring stronger held-out performance within the observed benchmark.

This comparison was included as an exploratory descriptive analysis of configuration-space behaviour rather than as a formal model-selection framework. Because held-out test performance contributes to the comparison, these results should not be interpreted as an unbiased external validation of a configuration-selection rule. Instead, they provide a practical summary of how different heuristic choices behave within the current benchmark and help contextualise the extent to which training-only selection may favour overfit configurations.

\subsection{Association between heritability magnitude and downstream PRS performance}
To assess whether the magnitude of the heritability estimate was associated with downstream predictive performance, we computed Pearson and Spearman correlations between the fold-specific $h^2$ value and three outcomes across all 86 configurations: mean test AUC (or $R^2$ for body mass index), mean training AUC, and the train--test gap $\Delta$. Correlations were computed per phenotype per tool, pooled across all phenotypes, and separately per estimation method. These associations are interpreted descriptively as properties of the joint benchmark behaviour of estimation and PRS parameterisation rather than as proof that a given estimator is intrinsically valid or invalid.
 
To characterise the sensitivity of downstream PRS performance to heritability configuration choice, we computed, for each phenotype and tool, the standard deviation, range, interquartile range, and coefficient of variation of test performance across all 86 configurations. To assess whether the choice of heritability estimation method systematically affected downstream PRS test performance, we computed the mean test AUC produced by each of the ten estimation methods across phenotypes and applied a Friedman test across methods and phenotypes. 

%To find the robustness of PRS performance to heritability configuration choice, we identified the top 10 configurations per phenotype and tool using the delta-constrained rule and recorded the train AUC, test AUC, and $h^2$ value for each. Summary statistics (mean, SD, range) across these configurations were computed to quantify how much train and test performance varied when $h^2$ changed substantially. To compare GCTA-SBLUP and LDpred2-lassosum2 directly, we computed the mean test AUC per phenotype averaged across all 86 configurations and the pooled mean per model condition. Pairwise differences were tested using Wilcoxon signed-rank tests across phenotypes and Mann--Whitney U tests pooled across all configurations.

%Post-hoc mean ranks were computed by ranking methods within each phenotype and averaging across phenotypes, where a lower mean rank indicates better average performance within the benchmark.

\section{Results}
\subsection{Overview of heritability-estimation results}
\begin{figure*}[!ht]
    \centering
    \includegraphics[width=0.9\linewidth]{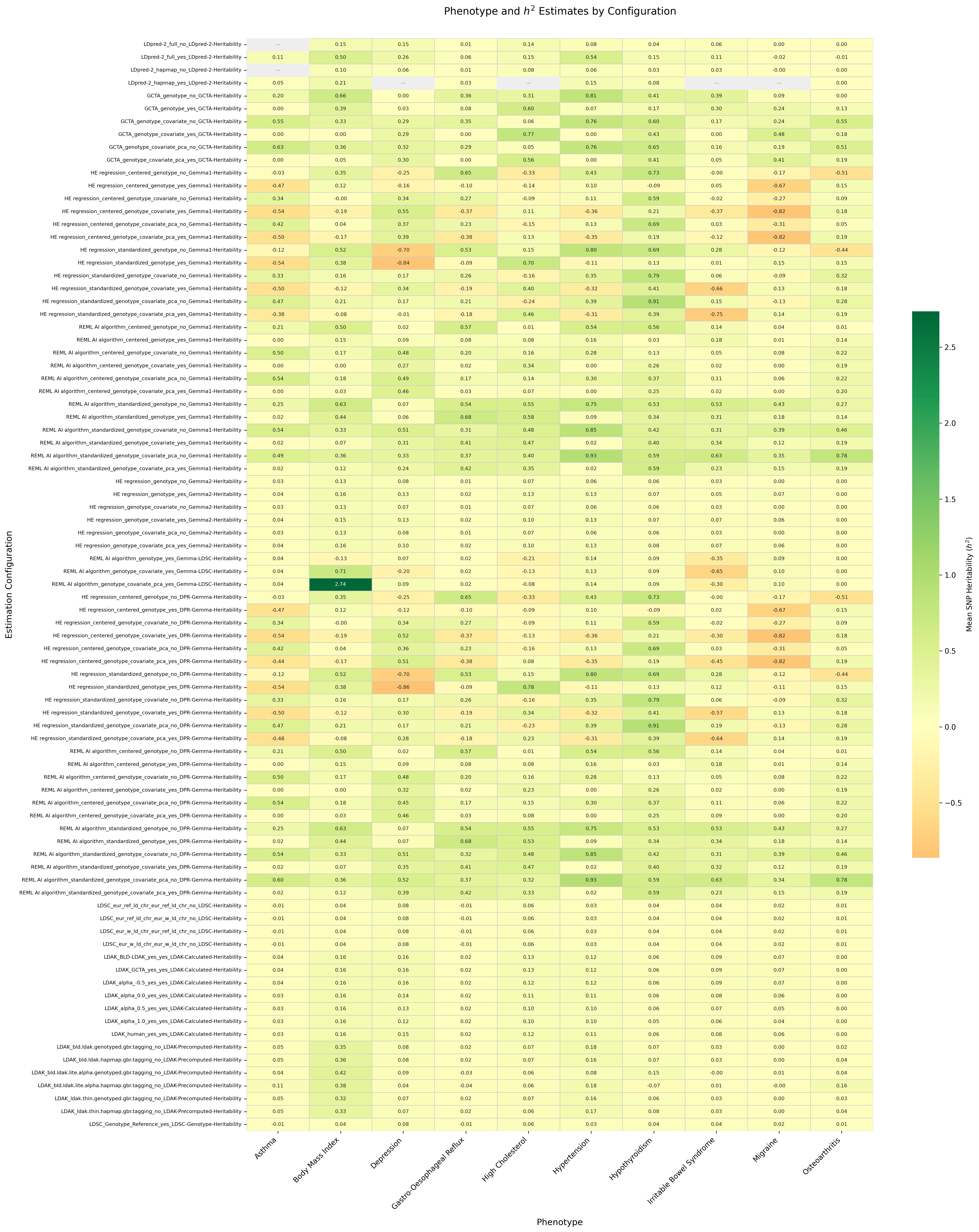}
    \caption{\textbf{Heatmap of mean SNP heritability estimates across all 86 configurations and 10 phenotypes.} Each row corresponds to one unique estimation configuration defined by the tool, model variant, clumping and pruning setting, and number of PCA components included; each column corresponds to one phenotype. Cell values represent the mean $h^2$ averaged across five cross-validation folds for the training set. Grey cells (—) indicate configurations for which no valid estimate was obtained for that phenotype specifically for LDpred2 when the number of variants are less than 15 percent between GWAS and the genotype data in settings when all the snps not excluding Hapmap were used. Negative values are concentrated in unconstrained Haseman--Elston regression configurations and reflect finite-sample sampling variability rather than estimator failure. Fold-level standard errors and 95\% confidence intervals for all configurations are reported in Supplementary Table~S1 (Heritability Estimates).}
    \label{Heatmap}
\end{figure*}

We analysed 86 distinct heritability estimates generated across six tool families ten method groups, and 86 configurations for all 10 phenotypes (Table~\ref{Table2}). Figure \ref{Heatmap} presents the heatmap of the heritability values for all phenotypes and 86 distinct heritability configuration variants for all methods we considered.  

\subsection{Distribution and variability of heritability estimates}
\begin{figure*}[!ht]
    \centering
    \includegraphics[width=\linewidth]{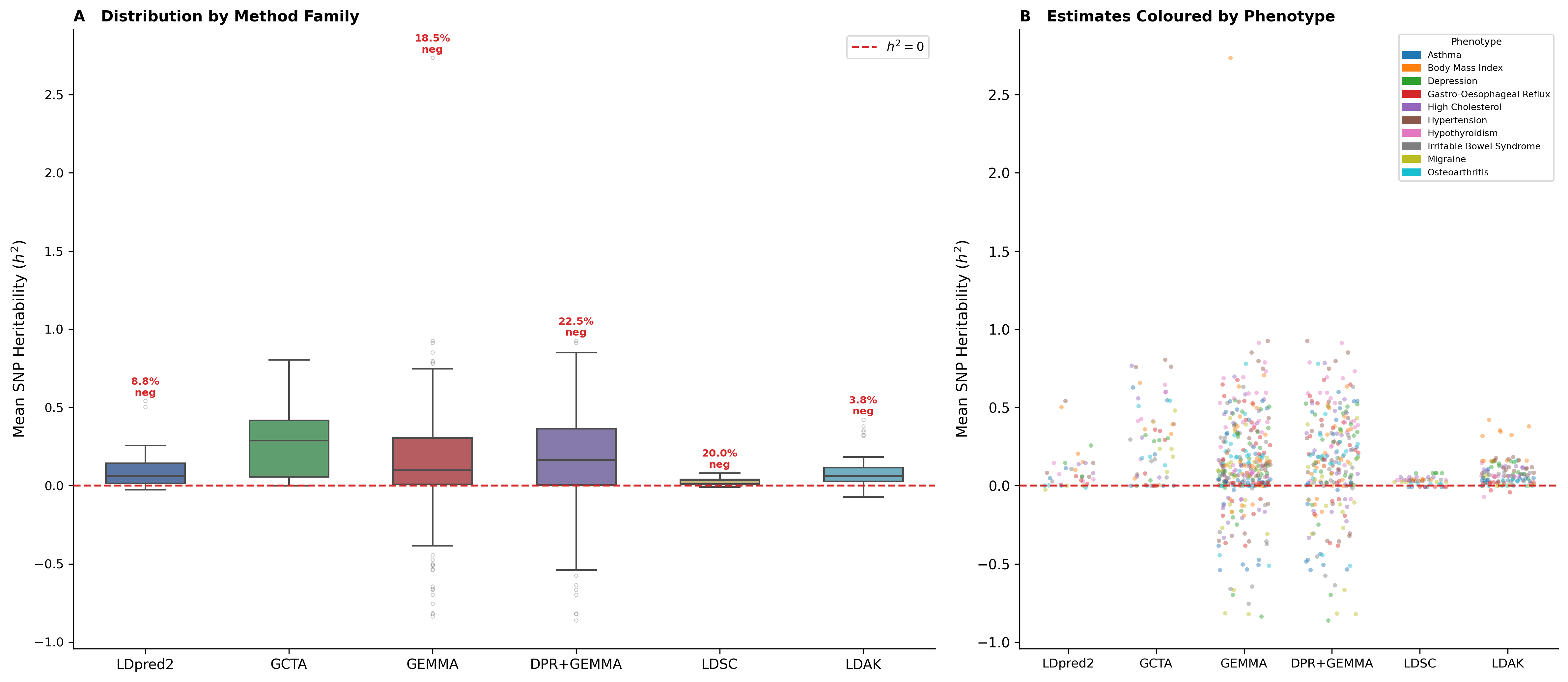}
    \caption{\textbf{Distribution of SNP heritability estimates across method families.} Each box summarises the mean $h^2$ values produced by one method family across all configurations and phenotypes. The central line is the median, box edges are the interquartile range, and whiskers extend to 1.5$\times$IQR; individual points beyond the whiskers are shown. The dashed red line marks $h^2 = 0$; estimates below this line reflect finite-sample sampling variability in unconstrained Haseman--Elston regression variants and are retained as benchmark outputs. Panel B shows the same estimates as a strip plot coloured by phenotype, illustrating which phenotypes drive extreme values within each family. }
    \label{fig:distribution}
\end{figure*}

Across all 844 configuration-level mean estimates, $h^2$ ranged from $-0.862$ to $2.735$ (mean $= 0.134$, SD $= 0.284$), reflecting substantial sensitivity to estimation strategy (Figure~\ref{fig:distribution}). At the method-family level, LDSC was the most stable estimator (SD $= 0.026$, mean $h^2 = 0.030$, CV $= 0.852$), followed by LDAK (SD $= 0.081$, CV $= 1.002$) and GCTA (SD $= 0.239$, CV $= 0.835$). DPR+GEMMA exhibited the highest variability (SD $= 0.328$, CV $= 2.115$), followed closely by GEMMA (SD $= 0.325$, CV $= 2.459$), both of which include Haseman--Elston regression variants that produced extreme values under low-signal conditions  regardless of the genotype matrix standardisation method or the inclusion of covariates and PCA. GCTA was the only family that produced no negative estimates (0.0\%), while DPR+GEMMA (22.5\%), GEMMA (18.5\%), and LDSC (20.0\%) produced the highest proportions of negative values. Negative estimates in the LDSC family arose from a small number of configurations under low-SNP conditions and should be distinguished from the structurally negative estimates produced by unconstrained HE regression in GEMMA and DPR+GEMMA.  

\subsection{Negative heritability estimates}
Negative heritability estimates were observed in 133 of 844 configuration-level estimates (15.4\% overall) and were not distributed uniformly across method families or estimator classes. Negative estimates were concentrated in DPR+GEMMA (22.5\%), LDSC (20.0\%), and GEMMA (18.5\%), while GCTA produced none (0.0\%) and LDAK very few (3.8\%). The occurrence of negative values was strongly associated with estimator type: mixed estimators combining HE regression and REML produced negative $h^2$ in 19.8\% of configurations, constrained REML-based estimators in 8.9\%, and unconstrained estimators in 8.4\% of configurations. At the method-family level, DPR+GEMMA produced the highest proportion of negative estimates (22.5\%, $n = 54$ of 240), followed by LDSC (20.0\%, $n = 10$ of 50) and GEMMA (18.5\%, $n = 61$ of 330), while GCTA produced none (0.0\%, $n = 0$ of 60) and LDAK produced very few (3.8\%, $n = 5$ of 130). This pattern is consistent with the mathematical properties of each estimator class: REML imposes a non-negativity constraint during optimisation, whereas HE regression and LD score regression do not, and therefore return negative solutions when the signal-to-noise ratio is low. Across phenotypes, Migraine showed the highest concentration of negative estimates while Hypothyroidism showed the fewest, consistent with differences in trait heritability and SNP coverage rather than systematic estimator bias. Negative estimates are retained in the benchmark as configuration outputs and are not treated as grounds for excluding a method or configuration.

\subsection{Stability and reliability of heritability estimates}
Fold-derived standard errors varied substantially across method families, reflecting differences in estimation consistency across cross-validation folds. LDSC was the most stable family (mean fold-SE $= 0.0000$, mean CV $= 0.000$), followed by LDAK (mean fold-SE $= 0.0004$, mean CV $= 0.007$) and LDpred2 (mean fold-SE $= 0.0065$, mean CV $= 0.062$). DPR+GEMMA exhibited the highest instability (mean fold-SE $= 0.131$, mean CV $= 1.714$), followed by GEMMA (mean fold-SE $= 0.118$, mean CV $= 1.391$), driven by HE regression variants that produced extreme estimates under low-signal conditions. The proportion of configurations yielding a statistically reliable positive estimate (fold-based 95\% CI entirely above zero) also varied substantially. LDAK produced the highest proportion of reliable positive estimates (96.2\%, $n = 125$ of 130), followed by LDpred2 (90.9\%, $n = 30$ of 33) and LDSC (80.0\%, $n = 40$ of 50), while DPR+GEMMA produced the lowest (45.4\%, $n = 109$ of 240). Across phenotypes, Hypothyroidism showed the highest overall reliability (89.4\%) and Gastro-Oesophageal Reflux the lowest (48.2\%), consistent with differences in trait heritability and SNP coverage. Even within phenotypes with high overall reliability, individual configurations varied in whether their CI excluded zero, underscoring the configuration-sensitivity of heritability estimation. Full fold-SE values and confidence intervals for all 86 configurations and 10 phenotypes are provided in Supplementary Table~S1 (Heritability Estimates).

\subsection{Effect of SNP count on heritability estimates}

Across all methods and phenotypes, the number of variants included in heritability estimation was positively correlated \cite{Yang2010,Gudbjartsson2008} with the resulting $h^2$ estimate (Spearman $r = 0.330$, $p < 0.001$), consistent with the expectation that broader SNP coverage captures additional additive genetic signal. At the method-family level, a statistically significant positive correlation was observed in four of six families: GCTA ($r = 0.447$, $p < 0.001$), GEMMA ($r = 0.329$, $p < 0.001$), DPR+GEMMA ($r = 0.360$, $p < 0.001$), and LDSC ($r = 0.515$, $p < 0.001$). LDpred2 ($r = 0.133$, $p = 0.455$) and LDAK ($r = 0.031$, $p = 0.723$) showed no significant association, likely because these methods use precomputed or fixed reference panels that decouple variant count from estimation variance. At the phenotype level, a significant positive correlation was observed in seven of ten phenotypes, with Hypertension ($r = 0.640$) and Hypothyroidism ($r = 0.548$) showing the strongest associations. High Cholesterol was the only phenotype showing a significant negative correlation ($r = -0.313$, $p = 0.004$), which may reflect the high baseline SNP overlap for that phenotype across configurations. These results indicate that variant filtering decisions represent a non-trivial source of variability in heritability estimation and should be reported explicitly alongside $h^2$ values.

\subsection{Effect of estimation hyperparameters on heritability}
To determine which analytical choices drove heritability variation, we tested eleven binary hyperparameter contrasts across all method families and phenotypes using Mann--Whitney U tests (Supplementary Table~S6); per-family results are reported in Supplementary Table~S7. Ten of eleven contrasts were statistically significant when pooled across all methods ($p < 0.05$). The largest effects were associated with algorithm choice and GRM construction. Configurations using REML produced substantially higher mean $h^2$ estimates than those using other algorithms (mean $h^2 = 0.252$ versus $0.079$, $\Delta h^2 = 0.173$, $p < 0.001$), while configurations using HE regression produced lower estimates (mean $h^2 = 0.043$ versus $0.185$, $\Delta h^2 = -0.142$, $p < 0.001$). A standardised GRM was associated with higher estimates than a centred GRM ($\Delta h^2 = 0.131$, $p < 0.001$). Clumping and pruning consistently reduced heritability estimates when pooled across all methods ($\Delta h^2 = -0.129$, $p < 0.001$), reflecting the reduction in variant count passed to each estimator. Covariate and PCA inclusion produced smaller but statistically significant positive effects ($\Delta h^2 = 0.053$, $p < 0.001$ and $\Delta h^2 = 0.053$, $p = 0.004$, respectively). Use of genotype data was associated with higher estimates than use of GWAS summary statistics alone ($\Delta h^2 = 0.083$, $p < 0.001$), and use of GWAS summary statistics was associated with lower estimates ($\Delta h^2 = -0.084$, $p < 0.001$). HapMap-based tagging produced a small but significant reduction relative to other tagging strategies ($\Delta h^2 = -0.062$, $p = 0.031$). Genotyped tagging was the only contrast that did not reach significance ($\Delta h^2 = -0.052$, $p = 0.127$). At the method-family level, the significance and direction of these effects were not uniform. Algorithm choice had the largest and most consistent within-family effect: in GEMMA, REML configurations produced mean $h^2 = 0.235$ versus $0.047$ for HE regression ($\Delta h^2 = 0.188$, $p < 0.001$), and in DPR+GEMMA, the corresponding difference was $0.274$ versus $0.037$ ($\Delta h^2 = 0.236$, $p < 0.001$). Clumping and pruning had a significant negative effect in GCTA ($\Delta h^2 = -0.163$, $p = 0.007$), GEMMA ($\Delta h^2 = -0.163$, $p < 0.001$), and DPR+GEMMA ($\Delta h^2 = -0.223$, $p < 0.001$), but not in LDpred2, LDSC, or LDAK, where variant selection is handled differently. Covariate and PCA inclusion produced non-significant effects within each individual family, indicating that their pooled significance was driven by between-family differences in baseline heritability rather than a consistent within-family effect. These findings confirm that reported heritability values should always be accompanied by the full specification of the estimation configuration, including algorithm, input data type, variant filtering, and covariate choices.

\subsection{Inter-method correlation and agreement}
\begin{figure*}
    \centering
    \includegraphics[width=1\linewidth]{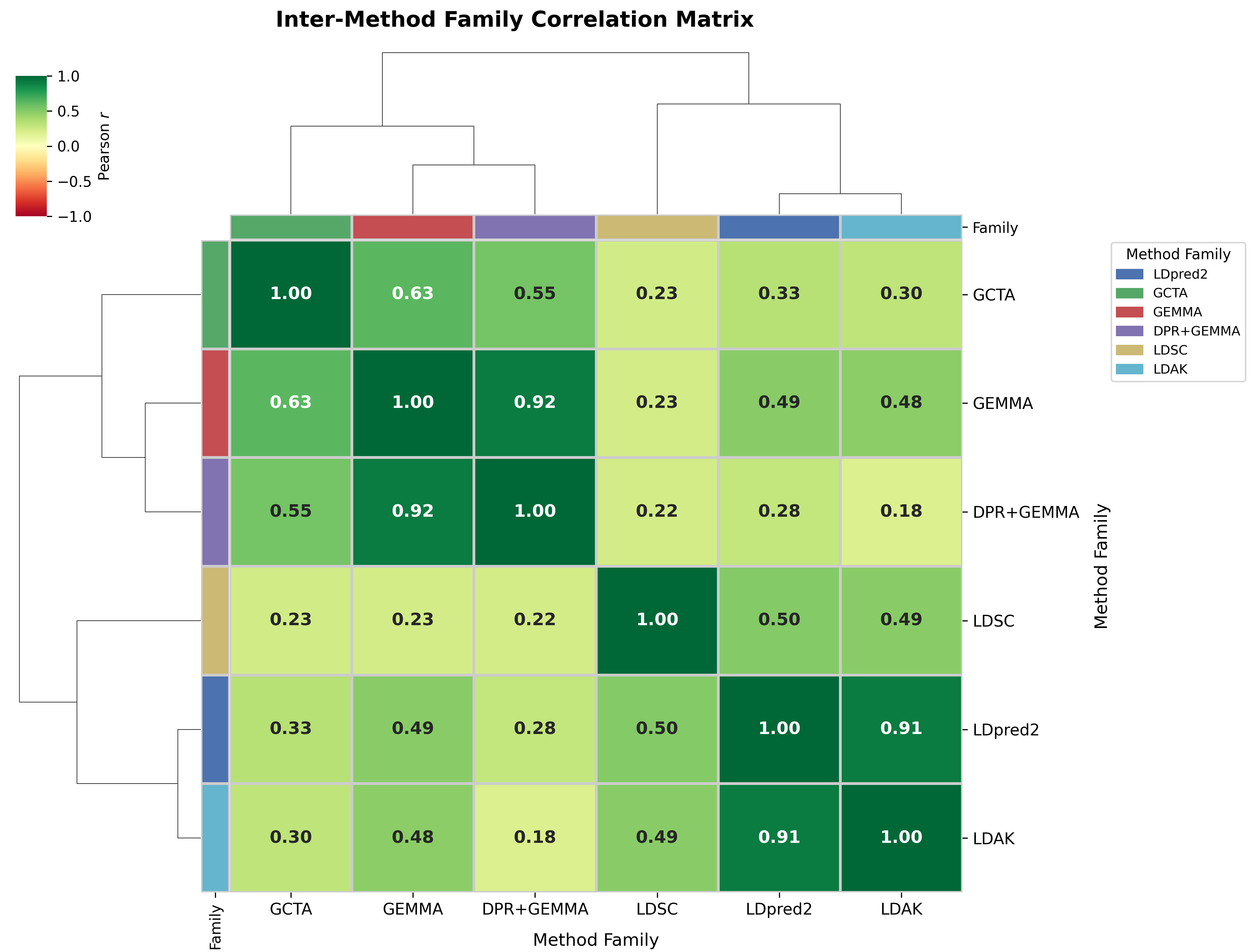}
    \caption{Inter-method family correlation matrix for heritability estimates. Pairwise Pearson correlations between all six method families are shown, computed over mean $h^2$ profiles across the ten phenotypes. Hierarchical clustering with average linkage and Euclidean distance was applied to the correlation values to produce the dendrogram. Cell values report the Pearson $r$; statistically significant pairs ($p < 0.05$) are indicated. Methods within the same family do not show significantly stronger agreement than cross-family pairs (Mann--Whitney $p = 0.459$), indicating that algorithmic class rather than software family membership drives inter-method agreement.}
    \label{fig:intercorr}
\end{figure*}

To assess whether methods agreed on the relative ranking of phenotypes by heritability, we computed pairwise Pearson correlations between all ten estimation methods over the ten phenotypes (Figure~\ref{fig:intercorr}). Of 45 method pairs, 13 showed a statistically significant positive correlation ($p < 0.05$), and all significant pairs were positively correlated. The most strongly correlated pair was LDSC-Genotype and LDSC-Heritability ($r = 1.000$, $p < 0.001$), which share the same underlying algorithm applied to different input resources. The second strongest correlation was between DPR-Gemma and Gemma1-Heritability ($r = 0.993$, $p < 0.001$), both of which use HE regression and REML applied to GEMMA-derived relatedness matrices. Strong cross-family correlations were also observed: Gemma2-Heritability and LDAK-Calculated-Heritability ($r = 0.958$, $p < 0.001$), Gemma2-Heritability and LDpred2 ($r = 0.927$, $p < 0.001$), and LDAK-Calculated and LDSC ($r = 0.806$, $p < 0.001$). The weakest correlations were between Gemma-LDSC and LDSC-Heritability ($r = 0.010$, $p = 0.978$) and between Gemma-LDSC and LDSC-Genotype ($r = 0.010$, $p = 0.978$), indicating that these methods disagreed substantially on the relative ordering of phenotypes. At the method-family level, GEMMA and DPR+GEMMA showed the strongest family-level agreement ($r = 0.918$, $p < 0.001$), while GCTA and LDSC showed the weakest ($r = 0.231$, $p = 0.522$). Methods from the same family did not show significantly higher pairwise correlations than methods from different families (mean same-family $r = 0.536$ versus different-family $r = 0.432$; Mann--Whitney $p = 0.459$), indicating that algorithmic similarity rather than software family membership drove inter-method agreement.  

\subsection{Matched-input controlled comparison}

To isolate estimator-level differences from preprocessing effects, we defined a matched-input subset in which each method and phenotype was represented by the configuration retaining the largest number of variants, with no clumping preferred as a tiebreaker. Under these matched-input conditions, heritability estimates still varied across methods (Kruskal--Wallis $H = 14.14$, $p = 0.117$), with mean $h^2$ ranging from $-0.023$ (Gemma-LDSC-Heritability) to $0.323$ (GCTA-Heritability). Although the between-method difference was not statistically significant at the global level, the spread of matched-input estimates indicates that variability across the full benchmark cannot be explained solely by variant filtering decisions. Rather, the observed differences are also consistent with estimator-level differences in modelling assumptions, LD handling, and variance component algorithms.

Same-family method pairs did not show significantly stronger agreement than different-family pairs (mean same-family $r = 0.395$ versus different-family $r = 0.292$; Mann--Whitney $p = 0.691$), further suggesting that shared software family membership alone does not determine agreement once input conditions are controlled. Taken together, these matched-input analyses support the interpretation that the benchmark captures both preprocessing-driven and estimator-driven sources of heritability variability.

\subsection{Baseline PRS performance}
Null model performance varied substantially across phenotypes. High Cholesterol showed the highest null model test AUC (0.852, 95\% CI 0.848--0.856), followed by Hypertension (0.745, 95\% CI 0.737--0.752) and Gastro-Oesophageal Reflux (0.690, 95\% CI 0.683--0.696). Asthma and Migraine showed the lowest null model test AUCs (0.520 and 0.526 respectively), indicating that covariates alone provided limited discriminative signal for these traits. Null model performance was identical between GCTA-SBLUP and LDpred2-lassosum2 for all phenotypes, as expected given that the null model does not depend on the PRS tool. The PRS-only model performed below the null model for most phenotypes under GCTA-SBLUP, with the exceptions of Depression (test AUC 0.580) and Irritable Bowel Syndrome (0.618), which marginally exceeded their respective null AUCs. Under LDpred2-lassosum2, PRS-only performance was consistently lower than under GCTA-SBLUP, with particularly large deficits for High Cholesterol (test AUC 0.351 versus 0.755) and Gastro-Oesophageal Reflux (0.391 versus 0.537). Body mass index showed strongly negative PRS-only $R^2$ values for both tools (GCTA-SBLUP: $-28.849$; LDpred2-lassosum2: $-28.716$), reflecting a known artefact of applying raw shrinkage-based scores to a continuous phenotype without standardisation. The full model test performance closely matched or fell below the null model for most phenotypes, indicating that the PRS contributed minimal additional discriminative signal above the covariate baseline on average across configurations. Full per-phenotype baseline performance metrics are provided in Supplementary Table~S2.

\subsection{Exploratory comparison of configuration selection heuristics}
The delta-constrained heuristic produced higher observed test performance and smaller train--test gaps than both best-train and random selection across the current benchmark (Supplementary Table~S3). Compared with random selection, the delta heuristic achieved higher mean test AUC for both GCTA-SBLUP ($\Delta = +0.106$, Wilcoxon $p = 0.014$) and LDpred2-lassosum2 ($\Delta = +0.102$, $p = 0.004$), together with smaller train--test gaps for both tools ($p = 0.002$). Relative to best-train selection, the observed gains were larger: mean test AUC increased by $+0.253$ for GCTA-SBLUP ($p = 0.004$) and $+0.084$ for LDpred2-lassosum2 ($p = 0.010$), while the train--test gap was reduced by up to 0.336. These results indicate that training-only selection tends to favour more overfit configurations within this benchmark, whereas the delta-constrained heuristic preferentially identifies configurations with a more favourable balance between apparent fit and held-out behaviour. However, because test performance contributes to this comparison, these findings should be interpreted as descriptive and exploratory rather than as a formal external validation of a model-selection procedure. The oracle comparison is included only as a benchmark upper bound for context and not as a deployable selection strategy.

\subsection{Association between heritability estimates and PRS performance}

Pearson and Spearman correlations between fold-specific $h^2$ values and PRS test performance are reported in Table~\ref{tab:h2_prs_corr}. Pooled across all phenotypes, the association between $h^2$ magnitude and test performance was negligible for both GCTA-SBLUP (Pearson $r = -0.023$, $p = 0.188$) and LDpred2-lassosum2 (Pearson $r = +0.014$, $p = 0.427$), confirming that $h^2$ magnitude was not a reliable predictor of downstream PRS accuracy across the full configuration space. Substantial heterogeneity was observed across phenotypes and estimation methods. For Depression, higher $h^2$ was positively associated with test performance under both GCTA-SBLUP (Pearson $r = +0.309$, $p < 0.001$) and LDpred2-lassosum2 (Pearson $r = +0.292$, $p < 0.001$), while for Asthma the association was negative under LDpred2-lassosum2 ($r = -0.232$, $p < 0.001$) and for Gastro-Oesophageal Reflux it was negative under both tools (GCTA-SBLUP: $r = -0.160$, $p = 0.005$; LDpred2-lassosum2: $r = -0.160$, $p = 0.005$). Hypothyroidism and High Cholesterol showed no significant association under either tool, indicating that PRS performance for these traits was entirely insensitive to heritability configuration choice. The correlation with the train--test gap was negative and significant for Depression under both tools (GCTA-SBLUP: $r = -0.342$, $p < 0.001$; LDpred2-lassosum2: $r = -0.231$, $p < 0.001$), indicating that higher $h^2$ reduced overfitting for this trait, whereas for Gastro-Oesophageal Reflux the relationship was reversed ($r = +0.145$, $p = 0.010$). At the method level, LDAK-Precomputed-Heritability showed the strongest negative association with test performance under both GCTA-SBLUP (Pearson $r = -0.718$, $p < 0.001$) and LDpred2-lassosum2 ($r = -0.727$, $p < 0.001$), indicating that the $h^2$ values returned by this method consistently miscalibrated PRS shrinkage. By contrast, Gemma1 and DPR-Gemma showed weak positive associations under LDpred2-lassosum2 ($r \approx +0.088$, $p < 0.01$), and GCTA-Heritability, LDSC, and LDSC-Genotype showed near-zero correlations under both tools.

\begin{table*}[!ht]
\centering
\caption{\textbf{Pearson and Spearman correlations between fold-specific $h^2$ and PRS performance.} Pearson $r$ (Test) quantifies the linear association between $h^2$ and mean test AUC or $R^2$; Pearson $r$ (Train) with training performance; Pearson $r$ ($\Delta$) with the train--test gap. Spearman $r$ (Test) is the rank-based equivalent. *** $p < 0.001$; ** $p < 0.01$; * $p < 0.05$; ns = not significant.}
\label{tab:h2_prs_corr}
\resizebox{\textwidth}{!}{%
\begin{tabular}{llrrrrrrrr}
\toprule
\multirow{2}{*}{\textbf{Phenotype}} & \multirow{2}{*}{\textbf{Tool}} & \multirow{2}{*}{\textbf{N}} &
\multicolumn{2}{c}{\textbf{Pearson $r$ (Test)}} &
\multicolumn{2}{c}{\textbf{Pearson $r$ (Train)}} &
\multicolumn{2}{c}{\textbf{Pearson $r$ ($\Delta$)}} &
\textbf{Spearman $r$ (Test)} \\
\cmidrule(lr){4-5}\cmidrule(lr){6-7}\cmidrule(lr){8-9}
& & & $r$ & Sig & $r$ & Sig & $r$ & Sig & $r$ (Sig) \\
\midrule
\multirow{2}{*}{Asthma}
 & GCTA-SBLUP & 313 & $-$0.1014 & ns  & $-$0.0617 & ns  & $+$0.0673 & ns  & $-$0.0802 (ns) \\
 & LDpred2    & 313 & $-$0.2318 & *** & $-$0.2262 & *** & $+$0.0359 & ns  & $-$0.1217 (*) \\
\midrule
\multirow{2}{*}{Body Mass Index}
 & GCTA-SBLUP & 319 & $+$0.1319 & *   & $+$0.0023 & ns  & $-$0.1320 & *   & $+$0.1495 (**) \\
 & LDpred2    & 363 & $+$0.1554 & **  & $+$0.0316 & ns  & $-$0.1551 & **  & $+$0.1695 (**) \\
\midrule
\multirow{2}{*}{Depression}
 & GCTA-SBLUP & 356 & $+$0.3085 & *** & $-$0.1819 & *** & $-$0.3419 & *** & $+$0.1464 (**) \\
 & LDpred2    & 336 & $+$0.2923 & *** & $-$0.0048 & ns  & $-$0.2312 & *** & $+$0.1663 (**) \\
\midrule
\multirow{2}{*}{Gastro-Oesophageal Reflux}
 & GCTA-SBLUP & 314 & $-$0.1600 & **  & $-$0.0592 & ns  & $+$0.1449 & *   & $-$0.1164 (*) \\
 & LDpred2    & 314 & $-$0.1599 & **  & $-$0.0565 & ns  & $+$0.1453 & **  & $-$0.1011 (ns) \\
\midrule
\multirow{2}{*}{High Cholesterol}
 & GCTA-SBLUP & 342 & $+$0.0144 & ns  & $+$0.2268 & *** & $-$0.0087 & ns  & $+$0.0460 (ns) \\
 & LDpred2    & 322 & $+$0.0148 & ns  & $+$0.2356 & *** & $-$0.0089 & ns  & $+$0.0138 (ns) \\
\midrule
\multirow{2}{*}{Hypertension}
 & GCTA-SBLUP & 359 & $+$0.1463 & **  & $+$0.0678 & ns  & $-$0.1184 & *   & $+$0.1773 (***) \\
 & LDpred2    & 359 & $+$0.0991 & ns  & $-$0.0708 & ns  & $-$0.0939 & ns  & $+$0.1111 (*) \\
\midrule
\multirow{2}{*}{Hypothyroidism}
 & GCTA-SBLUP & 380 & $-$0.0577 & ns  & $-$0.0287 & ns  & $+$0.0257 & ns  & $+$0.0220 (ns) \\
 & LDpred2    & 380 & $-$0.0539 & ns  & $-$0.0063 & ns  & $+$0.0385 & ns  & $+$0.0154 (ns) \\
\midrule
\multirow{2}{*}{Irritable Bowel Syndrome}
 & GCTA-SBLUP & 337 & $+$0.0017 & ns  & $-$0.0902 & ns  & $-$0.1452 & **  & $+$0.0009 (ns) \\
 & LDpred2    & 317 & $+$0.1328 & *   & $-$0.0299 & ns  & $-$0.1066 & ns  & $+$0.1147 (*) \\
\midrule
\multirow{2}{*}{Migraine}
 & GCTA-SBLUP & 316 & $+$0.1038 & ns  & $-$0.0825 & ns  & $-$0.1294 & *   & $+$0.0606 (ns) \\
 & LDpred2    & 258 & $+$0.0539 & ns  & $-$0.1402 & *   & $-$0.0917 & ns  & $+$0.0437 (ns) \\
\midrule
\multirow{2}{*}{Osteoarthritis}
 & GCTA-SBLUP & 381 & $+$0.1673 & **  & $+$0.2522 & *** & $-$0.1076 & *   & $+$0.0304 (ns) \\
 & LDpred2    & 365 & $+$0.0988 & ns  & $+$0.0731 & ns  & $-$0.0859 & ns  & $+$0.0298 (ns) \\
\midrule
\multirow{2}{*}{\textbf{All phenotypes (pooled)}}
 & GCTA-SBLUP & 3417 & $-$0.0225 & ns  & $-$0.0764 & *** & $+$0.0025 & ns  & $+$0.0214 (ns) \\
 & LDpred2    & 3327 & $+$0.0138 & ns  & $-$0.0318 & ns  & $-$0.0273 & ns  & $+$0.0548 (**) \\
\bottomrule
\end{tabular}%
}
\end{table*}

\subsection{Descriptive configuration summaries across phenotypes}
DPR-Gemma-Heritability was the most frequently represented method under the delta-constrained rule, appearing repeatedly across phenotypes under both PRS frameworks despite producing some of the highest fold-level variability in heritability estimates. Associated $h^2$ values ranged from 0.0008 to 0.981, confirming that no single heritability magnitude was uniformly associated with favourable downstream behaviour. Test AUC and explained variance under these representative configurations ranged from $-0.038$ (Body Mass Index under GCTA-SBLUP) to 0.882 (High Cholesterol under GCTA-SBLUP). GCTA-SBLUP and LDpred2-lassosum2 showed similar held-out performance when summarised across phenotypes (GCTA mean $= 0.605$, LDpred2 mean $= 0.597$, Wilcoxon $p = 0.232$, ns). Full configuration details are provided in Supplementary Table~S4.

\subsection{Configuration-space behaviour and method ranking}
PRS performance was broadly insensitive to heritability configuration choice across binary phenotypes. Excluding Body Mass Index, test AUC standard deviations across all 86 configurations ranged from 0.034 (High Cholesterol, both tools) to 0.067 (Depression and Hypothyroidism), indicating that a change in heritability configuration typically moved test AUC by less than 0.07 units. Body Mass Index was the clear exception, with test $R^2$ standard deviations of 0.672 (GCTA-SBLUP) and 0.717 (LDpred2-lassosum2) and a range exceeding 1.58, driven by extreme miscalibration of PRS shrinkage when $h^2$ was overestimated for this continuous trait. A Friedman test did not detect a statistically significant difference in test AUC across the ten heritability estimation methods for either GCTA-SBLUP ($\chi^2 = 15.646$, $p = 0.075$) or LDpred2-lassosum2 ($\chi^2 = 15.109$, $p = 0.088$), confirming that no single estimation method consistently outperformed others in downstream PRS accuracy. 
%Under GCTA-SBLUP, DPR-Gemma-Heritability received the best mean rank (2.88) and LDSC-Heritability the worst (7.19); under LDpred2-lassosum2, Gemma1-Heritability ranked first (2.75) and GCTA-Heritability last (6.88).

Examination of the top 10 configurations per phenotype further supports the conclusion that $h^2$ magnitude alone does not determine downstream PRS behaviour. For High Cholesterol under GCTA-SBLUP, test AUC varied by only 0.0003 while $h^2$ ranged from 0.000 to 0.162. For Hypertension under LDpred2-lassosum2, train and test AUC were nearly unchanged across the top 10 configurations despite an $h^2$ range of 1.000. Full results are provided in Supplementary Table S5. Averaged across all 86 configurations, GCTA-SBLUP and LDpred2-lassosum2 showed near-identical mean test performance (GCTA mean $= 0.497$, LDpred2 mean $= 0.498$, Wilcoxon $p = 1.000$, ns). Within the exploratory delta-constrained ranking, GCTA-SBLUP showed higher observed performance than LDpred2-lassosum2 for eight of ten phenotypes (GCTA mean $= 0.605$, LDpred2 mean $= 0.597$, Wilcoxon $p = 0.232$, ns), whereas LDpred2-lassosum2 was higher for Hypothyroidism and Irritable Bowel Syndrome. Taken together, these results indicate that downstream PRS behaviour is only weakly coupled to heritability magnitude and that favourable configurations are phenotype-specific within the present benchmark.

\begin{table*}[!ht]
\centering
\caption{\textbf{Summary of top 10 configurations per phenotype.} Train SD and Test SD are the standard deviations of train and test AUC across the top 10 configurations selected by the delta-constrained rule. $h^2$ range is the difference between the maximum and minimum $h^2$ values among the top 10. A narrow Test SD with a wide $h^2$ range indicates that PRS performance was robust to heritability magnitude within the top-performing configurations.}
\label{tab:top10_summary}
\resizebox{\textwidth}{!}{%
\begin{tabular}{llrrrrrrr}
\toprule
\textbf{Phenotype} & \textbf{Tool} &
\textbf{Train mean} & \textbf{Train SD} &
\textbf{Test mean} & \textbf{Test SD} &
\textbf{Test range} &
\textbf{$h^2$ mean} & \textbf{$h^2$ range} \\
\midrule
\multirow{2}{*}{Asthma}
 & GCTA-SBLUP & 0.904 & 0.001 & 0.562 & 0.001 & 0.005 & 0.027 & 0.036 \\
 & LDpred2    & 0.897 & 0.017 & 0.538 & 0.019 & 0.049 & 0.195 & 0.538 \\
\midrule
\multirow{2}{*}{Body Mass Index}
 & GCTA-SBLUP & 0.555 & 0.000 & $-$0.038 & 0.000 & 0.000 & 0.474 & 0.170 \\
 & LDpred2    & 0.564 & 0.000 & $-$0.046 & 0.000 & 0.000 & 0.279 & 0.495 \\
\midrule
\multirow{2}{*}{Depression}
 & GCTA-SBLUP & 0.902 & 0.001 & 0.637 & 0.001 & 0.002 & 0.793 & 0.186 \\
 & LDpred2    & 0.744 & 0.000 & 0.599 & 0.000 & 0.000 & 0.186 & 0.750 \\
\midrule
\multirow{2}{*}{Gastro-Oesophageal Reflux}
 & GCTA-SBLUP & 0.981 & 0.000 & 0.773 & 0.000 & 0.000 & 0.095 & 0.396 \\
 & LDpred2    & 0.981 & 0.000 & 0.773 & 0.000 & 0.000 & 0.219 & 0.517 \\
\midrule
\multirow{2}{*}{High Cholesterol}
 & GCTA-SBLUP & 1.000 & 0.000 & 0.882 & 0.000 & 0.000 & 0.055 & 0.162 \\
 & LDpred2    & 1.000 & 0.000 & 0.881 & 0.000 & 0.000 & 0.236 & 0.907 \\
\midrule
\multirow{2}{*}{Hypertension}
 & GCTA-SBLUP & 0.956 & 0.003 & 0.863 & 0.001 & 0.003 & 0.054 & 0.531 \\
 & LDpred2    & 0.940 & 0.000 & 0.832 & 0.000 & 0.000 & 0.238 & 1.000 \\
\midrule
\multirow{2}{*}{Hypothyroidism}
 & GCTA-SBLUP & 0.772 & 0.034 & 0.627 & 0.013 & 0.053 & 0.354 & 0.964 \\
 & LDpred2    & 0.857 & 0.028 & 0.653 & 0.009 & 0.020 & 0.357 & 0.981 \\
\midrule
\multirow{2}{*}{Irritable Bowel Syndrome}
 & GCTA-SBLUP & 0.705 & 0.027 & 0.535 & 0.020 & 0.076 & 0.138 & 0.527 \\
 & LDpred2    & 0.732 & 0.000 & 0.584 & 0.000 & 0.000 & 0.425 & 0.838 \\
\midrule
\multirow{2}{*}{Migraine}
 & GCTA-SBLUP & 0.993 & 0.000 & 0.580 & 0.000 & 0.000 & 0.279 & 0.957 \\
 & LDpred2    & 0.993 & 0.000 & 0.580 & 0.000 & 0.000 & 0.146 & 0.980 \\
\midrule
\multirow{2}{*}{Osteoarthritis}
 & GCTA-SBLUP & 0.956 & 0.002 & 0.621 & 0.002 & 0.009 & 0.002 & 0.011 \\
 & LDpred2    & 0.938 & 0.000 & 0.580 & 0.000 & 0.000 & 0.618 & 0.999 \\
\bottomrule
\end{tabular}%
}
\end{table*}

\section{Discussion}

This study systematically benchmarked 86 heritability estimation configurations spanning six tool families and ten method groups across 10 UK Biobank phenotypes, and linked the resulting $h^2$ estimates to downstream PRS performance using GCTA-SBLUP and LDpred2-lassosum2. Three main findings emerge.

\textbf{First, SNP heritability is highly configuration-sensitive.} Across 844 configuration-level estimates, $h^2$ ranged from $-0.862$ to $2.735$ (mean $= 0.134$, SD $= 0.284$), and ten of eleven analytical contrasts significantly affected the returned value. Algorithm choice was the strongest driver, with REML-based configurations returning substantially higher estimates than alternative estimators, while HE-regression-based configurations returned lower estimates overall. GRM standardisation and clumping/pruning also had large effects, indicating that the reported value of $h^2$ depends not only on the software family but also on the full estimation regime, including estimator class, SNP set, GRM construction, and preprocessing choices. Matched-input comparisons showed that variability persisted even when methods were compared under maximal variant-retention conditions, suggesting that the estimator itself contributes materially to the observed differences, although preprocessing decisions remain an important source of variation across the full benchmark.

\textbf{Second, negative heritability estimates are primarily a property of unconstrained estimators rather than evidence of estimator failure.} Of 844 estimates, 133 (15.8\%) were negative and were concentrated in DPR+GEMMA, GEMMA, and LDSC-derived configurations, whereas GCTA produced no negative estimates. This pattern is consistent with the mathematical properties of the underlying estimators: constrained REML-based methods enforce non-negativity, whereas unconstrained HE-regression and LD-score-regression variants may return negative solutions under low signal-to-noise conditions. Negative estimates should therefore be interpreted as informative outputs of a given estimation regime under the conditions studied, rather than as sufficient evidence that a method is unsuitable. In this context, fold-based confidence intervals provide a more principled measure of reliability than the sign of the point estimate alone.

\textbf{Third, substantial upstream variability in heritability did not translate into strong downstream PRS instability.} Pooled across phenotypes, the Pearson correlation between $h^2$ and PRS test AUC was negligible for both GCTA-SBLUP ($r = -0.023$, $p = 0.188$) and LDpred2-lassosum2 ($r = +0.014$, $p = 0.427$), indicating that heritability magnitude alone was not a reliable predictor of downstream predictive accuracy. Across many phenotypes, markedly different $h^2$ values produced very similar test performance. Sensitivity analyses showed that, for the binary traits, test AUC variation across the 86 configurations was generally modest, reinforcing the conclusion that moderate changes in the heritability input do not systematically destabilise downstream PRS behaviour. At the method level, no single heritability estimation family consistently dominated downstream PRS performance across phenotypes, indicating that the relationship between $h^2$ and downstream prediction is context dependent rather than universal.

The practical implication is that heritability should be treated as a configuration-sensitive modelling parameter rather than as a universally stable scalar input. For applied PRS workflows, reporting a single $h^2$ value without the associated estimation specification may be misleading, because materially different analytical regimes can yield markedly different values while producing similar downstream predictive performance. Heritability estimates should therefore be reported together with the estimation context, including estimator class, preprocessing decisions, SNP inclusion strategy, and reference resources.

This study also has important limitations. First, the benchmark compares heritability estimation configurations rather than a perfectly matched head-to-head comparison of software alone; although the matched-input analysis reduces this concern, heterogeneous inputs and preprocessing choices remain part of the broader benchmark design. Second, the downstream PRS analyses were conducted in a relatively small phenotype set drawn from UK Biobank, and performance estimates may therefore be sensitive to phenotype-specific sample size and case--control balance. Third, for some phenotypes, the external GWAS summary statistics may overlap partially with UK Biobank, which could inflate apparent downstream predictive performance. Fourth, the exploratory comparison of configuration-selection heuristics includes held-out test performance and should therefore not be interpreted as a formal external validation of a deployable model-selection procedure. Rather, it serves as a descriptive analysis showing how different heuristic choices behave within the present configuration space.

Taken together, these results support a narrower but robust conclusion: SNP heritability is strongly analysis-dependent in practice, whereas downstream PRS performance is often comparatively insensitive to moderate variation in the heritability input. Future benchmarking studies would benefit from larger phenotype panels, stricter matched-input comparisons, explicit overlap audits for discovery and target datasets, and fully nested validation designs for configuration selection. Even so, the present results already show that the practical meaning of a reported $h^2$ value depends critically on how that value was obtained.
\section{Conclusion}

We benchmarked 86 heritability estimation configurations across six tool families, ten method groups, and 10 UK Biobank phenotypes, and linked the resulting $h^2$ estimates to downstream PRS performance using GCTA-SBLUP and LDpred2-lassosum2. Three conclusions are supported by the data. First, SNP heritability is a configuration-sensitive parameter: across 844 estimates, $h^2$ ranged from $-0.862$ to $2.735$, and ten of eleven analytical contrasts significantly affected the returned value, with algorithm selection and GRM construction among the dominant drivers. Second, negative heritability values are primarily a property of unconstrained estimators under low-signal conditions rather than evidence of estimator failure. Third, downstream PRS performance was comparatively robust to this upstream variability: pooled correlations between $h^2$ and test AUC were negligible for both PRS frameworks, and no single heritability estimation family consistently outperformed others in downstream accuracy across phenotypes.

Taken together, these findings show that SNP heritability should be interpreted as a configuration-sensitive modelling parameter rather than as a universally stable scalar quantity. Heritability estimates should therefore be reported together with the full estimation specification, including estimator class, preprocessing choices, SNP inclusion strategy, and reference resources. Within the present benchmark, exploratory configuration comparisons further showed that heuristics balancing apparent fit against train--test discrepancy can identify more stable observed behaviour than training-only selection, but formal validation of configuration-selection rules will require fully nested evaluation in future work. More broadly, these findings have direct implications for translational bioinformatics workflows where PRS are increasingly deployed as clinical risk stratification tools: the robustness of downstream prediction to heritability configuration choice supports the practical utility of PRS even when the optimal estimation strategy is uncertain, provided that configuration choices are transparently reported.

\clearpage

\section*{Funding}
M.M. is supported by a University of Queensland Research Training Program (RTP) Scholarship. D.B.A. is supported by an NHMRC Investigator Grant (GRNT2041888). Research supported by the NVIDIA Academic Grant Program.

\section*{Ethics statement}
This study used existing UK Biobank data accessed under application ID 50000. The authors confirm that ethics approval was not required for this secondary analysis of existing data.

\section*{Declaration of competing interest}
The authors declare that they have no competing interests.

\section*{Author contributions}
Muhammad Muneeb: Conceptualization, Methodology, Software, Formal analysis, Data curation, Writing -- Original draft preparation. David B. Ascher: Conceptualization, Methodology, Supervision, Writing -- Review and editing.

\section*{Data and software availability}
The genotype and phenotype data used in this study were accessed through the UK Biobank under application ID 50000 (\url{https://www.ukbiobank.ac.uk/}) and are subject to UK Biobank access restrictions; researchers may apply for access through the UK Biobank Access Management System. GWAS summary statistics were obtained from the GWAS Catalog (\url{https://www.ebi.ac.uk/gwas/}). Analysis code for this study is available at \url{https://github.com/MuhammadMuneeb007/Benchmarking-Heritability-Estimation-Strategies-Across-86-Configurations-and-Their-Downstream-Effect}. The following software tools were used: GEMMA (\url{https://github.com/genetics-statistics/GEMMA}), GCTA (\url{http://cnsgenomics.com/software/gcta/}), LDAK (\url{http://dougspeed.com/ldak/}), DPR (\url{https://github.com/biostatpzeng/DPR}), LDSC (\url{https://github.com/bulik/ldsc}), and SumHer (\url{http://dougspeed.com/sumher/}).

\section*{Declaration of generative AI and AI-assisted technologies in the manuscript preparation process}
During the preparation of this work the authors used Claude (Anthropic) and ChatGPT (OpenAI) in order to assist with manuscript writing, editing, and language improvement. After using these tools, the authors reviewed and edited the content as needed and take full responsibility for the content of the published article.

\section*{Supplementary Material}
Supplementary Table S1 reports, for each of the 86 estimation configurations and 10 phenotypes, the mean SNP heritability ($h^2$) averaged across five cross-validation folds, the fold-derived standard error (fold-SE), the 95\% confidence interval based on fold variation, and where available the tool-reported standard error and its corresponding CI. The column \texttt{significant\_fold} indicates whether the lower bound of the fold-based CI exceeded zero, providing a configuration-level indicator of whether the heritability estimate is statistically distinguishable from zero given cross-validation variability. Negative mean $h^2$ values reflect finite-sample variability in unconstrained Haseman--Elston regression variants and are retained as benchmark outputs. Supplementary Table S2 reports full per-phenotype baseline PRS performance metrics for the null, PRS-only, and full model conditions averaged across all 86 configurations and five cross-validation folds. Supplementary Table S3 reports full per-phenotype configuration selection strategy results comparing the delta-constrained heuristic, best-train selection, random selection, and oracle upper bound. Supplementary Table S4 reports full per-phenotype representative high-performing configurations identified under the delta-constrained rule. Supplementary Table S5 reports top 10 configuration summary statistics per phenotype including train and test performance variability and $h^2$ range. Supplementary Table S6 reports the effect of all eleven binary hyperparameter contrasts on SNP heritability estimates pooled across all method families and phenotypes. Supplementary Table S7 reports the per-family breakdown of hyperparameter effects on heritability estimates across all method families.

\printcredits

\bibliographystyle{cas-model2-names}
\bibliography{cas-refs}

@article{Sakaue2021,
  title = {A cross-population atlas of genetic associations for 220 human phenotypes},
  volume = {53},
  ISSN = {1546-1718},
  url = {http://dx.doi.org/10.1038/s41588-021-00931-x},
  DOI = {10.1038/s41588-021-00931-x},
  number = {10},
  journal = {Nature Genetics},
  publisher = {Springer Science and Business Media LLC},
  author = {Sakaue,  Saori and Kanai,  Masahiro and Tanigawa,  Yosuke and Karjalainen,  Juha and Kurki,  Mitja and Koshiba,  Seizo and Narita,  Akira and Konuma,  Takahiro and Yamamoto,  Kenichi and Akiyama,  Masato and Ishigaki,  Kazuyoshi and Suzuki,  Akari and Suzuki,  Ken and Obara,  Wataru and Yamaji,  Ken and Takahashi,  Kazuhisa and Asai,  Satoshi and Takahashi,  Yasuo and Suzuki,  Takao and Shinozaki,  Nobuaki and Yamaguchi,  Hiroki and Minami,  Shiro and Murayama,  Shigeo and Yoshimori,  Kozo and Nagayama,  Satoshi and Obata,  Daisuke and Higashiyama,  Masahiko and Masumoto,  Akihide and Koretsune,  Yukihiro and Ito,  Kaoru and Terao,  Chikashi and Yamauchi,  Toshimasa and Komuro,  Issei and Kadowaki,  Takashi and Tamiya,  Gen and Yamamoto,  Masayuki and Nakamura,  Yusuke and Kubo,  Michiaki and Murakami,  Yoshinori and Yamamoto,  Kazuhiko and Kamatani,  Yoichiro and Palotie,  Aarno and Rivas,  Manuel A. and Daly,  Mark J. and Matsuda,  Koichi and Okada,  Yukinori},
  year = {2021},
  month = sep,
  pages = {1415–1424}
}

@article{Howard2018,
  title = {Genome-wide association study of depression phenotypes in UK Biobank identifies variants in excitatory synaptic pathways},
  volume = {9},
  ISSN = {2041-1723},
  url = {http://dx.doi.org/10.1038/s41467-018-03819-3},
  DOI = {10.1038/s41467-018-03819-3},
  number = {1},
  journal = {Nature Communications},
  publisher = {Springer Science and Business Media LLC},
  author = {Howard,  David M. and Adams,  Mark J. and Shirali,  Masoud and Clarke,  Toni-Kim and Marioni,  Riccardo E. and Davies,  Gail and Coleman,  Jonathan R. I. and Alloza,  Clara and Shen,  Xueyi and Barbu,  Miruna C. and Wigmore,  Eleanor M. and Gibson,  Jude and Agee,  Michelle and Alipanahi,  Babak and Auton,  Adam and Bell,  Robert K. and Bryc,  Katarzyna and Elson,  Sarah L. and Fontanillas,  Pierre and Furlotte,  Nicholas A. and Hinds,  David A. and Huber,  Karen E. and Kleinman,  Aaron and Litterman,  Nadia K. and McCreight,  Jennifer C. and McIntyre,  Matthew H. and Mountain,  Joanna L. and Noblin,  Elizabeth S. and Northover,  Carrie A. M. and Pitts,  Steven J. and Sathirapongsasuti,  J. Fah and Sazonova,  Olga V. and Shelton,  Janie F. and Shringarpure,  Suyash and Tian,  Chao and Tung,  Joyce Y. and Vacic,  Vladimir and Wilson,  Catherine H. and Hagenaars,  Saskia P. and Lewis,  Cathryn M. and Ward,  Joey and Smith,  Daniel J. and Sullivan,  Patrick F. and Haley,  Chris S. and Breen,  Gerome and Deary,  Ian J. and McIntosh,  Andrew M.},
  year = {2018},
  month = apr 
}

@article{Loh2018,
  title = {Mixed-model association for biobank-scale datasets},
  volume = {50},
  ISSN = {1546-1718},
  url = {http://dx.doi.org/10.1038/s41588-018-0144-6},
  DOI = {10.1038/s41588-018-0144-6},
  number = {7},
  journal = {Nature Genetics},
  publisher = {Springer Science and Business Media LLC},
  author = {Loh,  Po-Ru and Kichaev,  Gleb and Gazal,  Steven and Schoech,  Armin P. and Price,  Alkes L.},
  year = {2018},
  month = jun,
  pages = {906–908}
}

@article{GuindoMartnez2021,
  title = {The impact of non-additive genetic associations on age-related complex diseases},
  volume = {12},
  ISSN = {2041-1723},
  url = {http://dx.doi.org/10.1038/s41467-021-21952-4},
  DOI = {10.1038/s41467-021-21952-4},
  number = {1},
  journal = {Nature Communications},
  publisher = {Springer Science and Business Media LLC},
  author = {Guindo-Martínez,  Marta and Amela,  Ramon and Bonàs-Guarch,  Silvia and Puiggròs,  Montserrat and Salvoro,  Cecilia and Miguel-Escalada,  Irene and Carey,  Caitlin E. and Cole,  Joanne B. and R\"{u}eger,  Sina and Atkinson,  Elizabeth and Leong,  Aaron and Sanchez,  Friman and Ramon-Cortes,  Cristian and Ejarque,  Jorge and Palmer,  Duncan S. and Kurki,  Mitja and Aragam,  Krishna and Florez,  Jose C. and Badia,  Rosa M. and Mercader,  Josep M. and Torrents,  David},
  year = {2021},
  month = apr 
}

@article{Eijsbouts2021,
  title = {Genome-wide analysis of 53, 400 people with irritable bowel syndrome highlights shared genetic pathways with mood and anxiety disorders},
  volume = {53},
  ISSN = {1546-1718},
  url = {http://dx.doi.org/10.1038/s41588-021-00950-8},
  DOI = {10.1038/s41588-021-00950-8},
  number = {11},
  journal = {Nature Genetics},
  publisher = {Springer Science and Business Media LLC},
  author = {Eijsbouts,  Chris and Zheng,  Tenghao and Kennedy,  Nicholas A. and Bonfiglio,  Ferdinando and Anderson,  Carl A. and Moutsianas,  Loukas and Holliday,  Joanne and Shi,  Jingchunzi and Shringarpure,  Suyash and Agee,  Michelle and Aslibekyan,  Stella and Auton,  Adam and Bell,  Robert K. and Bryc,  Katarzyna and Clark,  Sarah K. and Elson,  Sarah L. and Fletez-Brant,  Kipper and Fontanillas,  Pierre and Furlotte,  Nicholas A. and Gandhi,  Pooja M. and Heilbron,  Karl and Hicks,  Barry and Hinds,  David A. and Huber,  Karen E. and Jewett,  Ethan M. and Jiang,  Yunxuan and Kleinman,  Aaron and Lin,  Keng-Han and Litterman,  Nadia K. and Luff,  Marie K. and McCreight,  Jey C. and McIntyre,  Matthew H. and McManus,  Kimberly F. and Mountain,  Joanna L. and Mozaffari,  Sahar V. and Nandakumar,  Priyanka and Noblin,  Elizabeth S. and Northover,  Carrie A. M. and O’Connell,  Jared and Petrakovitz,  Aaron A. and Pitts,  Steven J. and Poznik,  G. David and Sathirapongsasuti,  J. Fah and Shastri,  Anjali J. and Shelton,  Janie F. and Tian,  Chao and Tung,  Joyce Y. and Tunney,  Robert J. and Vacic,  Vladimir and Wang,  Xin and Zare,  Amir S. and Voda,  Alexandru-Ioan and Kashyap,  Purna and Chang,  Lin and Mayer,  Emeran and Heitkemper,  Margaret and Sayuk,  Gregory S. and Ringel-Kulka,  Tamar and Ringel,  Yehuda and Chey,  William D. and Eswaran,  Shanti and Merchant,  Juanita L. and Shulman,  Robert J. and Bujanda,  Luis and Garcia-Etxebarria,  Koldo and Dlugosz,  Aldona and Lindberg,  Greger and Schmidt,  Peter T. and Karling,  Pontus and Ohlsson,  Bodil and Walter,  Susanna and Faresj,  Ashild O. and Simren,  Magnus and Halfvarson,  Jonas and Portincasa,  Piero and Barbara,  Giovanni and Usai-Satta,  Paolo and Neri,  Matteo and Nardone,  Gerardo and Cuomo,  Rosario and Galeazzi,  Francesca and Bellini,  Massimo and Latiano,  Anna and Houghton,  Lesley and Jonkers,  Daisy and Kurilshikov,  Alexander and Weersma,  Rinse K. and Netea,  Mihai and Tesarz,  Jonas and Gauss,  Annika and Goebel-Stengel,  Miriam and Andresen,  Viola and Frieling,  Thomas and Pehl,  Christian and Schaefert,  Rainer and Niesler,  Beate and Lieb,  Wolfgang and Hanevik,  Kurt and Langeland,  Nina and Wensaas,  Knut-Arne and Litleskare,  Sverre and Gabrielsen,  Maiken E. and Thomas,  Laurent and Thijs,  Vincent and Lemmens,  Robin and Van Oudenhove,  Lukas and Wouters,  Mira and Farrugia,  Gianrico and Franke,  Andre and H\"{u}benthal,  Matthias and Abecasis,  Gon\c{c}alo and Zawistowski,  Matthew and Skogholt,  Anne Heidi and Ness-Jensen,  Eivind and Hveem,  Kristian and Esko,  Tõnu and Teder-Laving,  Maris and Zhernakova,  Alexandra and Camilleri,  Michael and Boeckxstaens,  Guy and Whorwell,  Peter J. and Spiller,  Robin and McVean,  Gil and D’Amato,  Mauro and Jostins,  Luke and Parkes,  Miles},
  year = {2021},
  month = nov,
  pages = {1543–1552}
}

@article{Jiang2021,
  title = {A generalized linear mixed model association tool for biobank-scale data},
  volume = {53},
  ISSN = {1546-1718},
  url = {http://dx.doi.org/10.1038/s41588-021-00954-4},
  DOI = {10.1038/s41588-021-00954-4},
  number = {11},
  journal = {Nature Genetics},
  publisher = {Springer Science and Business Media LLC},
  author = {Jiang,  Longda and Zheng,  Zhili and Fang,  Hailing and Yang,  Jian},
  year = {2021},
  month = nov,
  pages = {1616–1621}
}

@article{McDonald2022,
  title = {Novel genetic loci associated with osteoarthritis in multi-ancestry analyses in the Million Veteran Program and UK Biobank},
  volume = {54},
  ISSN = {1546-1718},
  url = {http://dx.doi.org/10.1038/s41588-022-01221-w},
  DOI = {10.1038/s41588-022-01221-w},
  number = {12},
  journal = {Nature Genetics},
  publisher = {Springer Science and Business Media LLC},
  author = {McDonald,  Merry-Lynn N. and Lakshman Kumar,  Preeti and Srinivasasainagendra,  Vinodh and Nair,  Ashwathy and Rocco,  Alison P. and Wilson,  Ava C. and Chiles,  Joe W. and Richman,  Joshua S. and Pinson,  Sarah A. and Dennis,  Richard A. and Jagadale,  Vivek and Brown,  Cynthia J. and Pyarajan,  Saiju and Tiwari,  Hemant K. and Bamman,  Marcas M. and Singh,  Jasvinder A.},
  year = {2022},
  month = nov,
  pages = {1816–1826}
}

@article{Chung2021,
  title = {Statistical models and computational tools for predicting complex traits and diseases},
  volume = {19},
  ISSN = {2234-0742},
  url = {http://dx.doi.org/10.5808/gi.21053},
  DOI = {10.5808/gi.21053},
  number = {4},
  journal = {Genomics and; Informatics},
  publisher = {Korea Genome Organization},
  author = {Chung,  Wonil},
  year = {2021},
  month = dec,
  pages = {e36}
}

@article{Li2023,
  title = {Accurate and efficient estimation of local heritability using summary statistics and the linkage disequilibrium matrix},
  volume = {14},
  ISSN = {2041-1723},
  url = {http://dx.doi.org/10.1038/s41467-023-43565-9},
  DOI = {10.1038/s41467-023-43565-9},
  number = {1},
  journal = {Nature Communications},
  publisher = {Springer Science and Business Media LLC},
  author = {Li,  Hui and Mazumder,  Rahul and Lin,  Xihong},
  year = {2023},
  month = dec 
}

@article{Yang2010,
  title = {Common SNPs explain a large proportion of the heritability for human height},
  volume = {42},
  ISSN = {1546-1718},
  url = {http://dx.doi.org/10.1038/ng.608},
  DOI = {10.1038/ng.608},
  number = {7},
  journal = {Nature Genetics},
  publisher = {Springer Science and Business Media LLC},
  author = {Yang,  Jian and Benyamin,  Beben and McEvoy,  Brian P and Gordon,  Scott and Henders,  Anjali K and Nyholt,  Dale R and Madden,  Pamela A and Heath,  Andrew C and Martin,  Nicholas G and Montgomery,  Grant W and Goddard,  Michael E and Visscher,  Peter M},
  year = {2010},
  month = jun,
  pages = {565–569}
}

@article{Gudbjartsson2008,
  title = {Many sequence variants affecting diversity of adult human height},
  volume = {40},
  ISSN = {1546-1718},
  url = {http://dx.doi.org/10.1038/ng.122},
  DOI = {10.1038/ng.122},
  number = {5},
  journal = {Nature Genetics},
  publisher = {Springer Science and Business Media LLC},
  author = {Gudbjartsson,  Daniel F and Walters,  G Bragi and Thorleifsson,  Gudmar and Stefansson,  Hreinn and Halldorsson,  Bjarni V and Zusmanovich,  Pasha and Sulem,  Patrick and Thorlacius,  Steinunn and Gylfason,  Arnaldur and Steinberg,  Stacy and Helgadottir,  Anna and Ingason,  Andres and Steinthorsdottir,  Valgerdur and Olafsdottir,  Elinborg J and Olafsdottir,  Gudridur H and Jonsson,  Thorvaldur and Borch-Johnsen,  Knut and Hansen,  Torben and Andersen,  Gitte and Jorgensen,  Torben and Pedersen,  Oluf and Aben,  Katja K and Witjes,  J Alfred and Swinkels,  Dorine W and Heijer,  Martin den and Franke,  Barbara and Verbeek,  Andre L M and Becker,  Diane M and Yanek,  Lisa R and Becker,  Lewis C and Tryggvadottir,  Laufey and Rafnar,  Thorunn and Gulcher,  Jeffrey and Kiemeney,  Lambertus A and Kong,  Augustine and Thorsteinsdottir,  Unnur and Stefansson,  Kari},
  year = {2008},
  month = apr,
  pages = {609–615}
}

@article{https://doi.org/10.48550/arxiv.1110.6019,
  doi = {10.48550/ARXIV.1110.6019},
  url = {https://arxiv.org/abs/1110.6019},
  author = {Guan,  Yongtao and Stephens,  Matthew},
  title = {Bayesian variable selection regression for genome-wide association studies and other large-scale problems},
  publisher = {arXiv},
  year = {2011},
  copyright = {arXiv.org perpetual,  non-exclusive license}
}

@article{Zhou2013,
  title = {Polygenic Modeling with Bayesian Sparse Linear Mixed Models},
  volume = {9},
  ISSN = {1553-7404},
  url = {http://dx.doi.org/10.1371/journal.pgen.1003264},
  DOI = {10.1371/journal.pgen.1003264},
  number = {2},
  journal = {PLoS Genetics},
  publisher = {Public Library of Science (PLoS)},
  author = {Zhou,  Xiang and Carbonetto,  Peter and Stephens,  Matthew},
  editor = {Visscher,  Peter M.},
  year = {2013},
  month = feb,
  pages = {e1003264}
}

@article{MARDIA1984,
  title = {Maximum likelihood estimation of models for residual covariance in spatial regression},
  volume = {71},
  ISSN = {1464-3510},
  url = {http://dx.doi.org/10.1093/biomet/71.1.135},
  DOI = {10.1093/biomet/71.1.135},
  number = {1},
  journal = {Biometrika},
  publisher = {Oxford University Press (OUP)},
  author = {MARDIA,  K. V. and MARSHALL,  R. J.},
  year = {1984},
  pages = {135–146}
}

@article{Hujoel2022,
  title = {Incorporating family history of disease improves polygenic risk scores in diverse populations},
  volume = {2},
  ISSN = {2666-979X},
  url = {http://dx.doi.org/10.1016/j.xgen.2022.100152},
  DOI = {10.1016/j.xgen.2022.100152},
  number = {7},
  journal = {Cell Genomics},
  publisher = {Elsevier BV},
  author = {Hujoel,  Margaux L.A. and Loh,  Po-Ru and Neale,  Benjamin M. and Price,  Alkes L.},
  year = {2022},
  month = jul,
  pages = {100152}
}

@misc{Lindsay2005,
  title = {Method of Moments},
  ISBN = {9780470011812},
  url = {http://dx.doi.org/10.1002/0470011815.b2a15089},
  DOI = {10.1002/0470011815.b2a15089},
  journal = {Encyclopedia of Biostatistics},
  publisher = {Wiley},
  author = {Lindsay,  Bruce G.},
  year = {2005},
  month = feb 
}

@article{Corbeil1976,
  title = {Restricted Maximum Likelihood (REML) Estimation of Variance Components in the Mixed Model},
  volume = {18},
  ISSN = {0040-1706},
  url = {http://dx.doi.org/10.2307/1267913},
  DOI = {10.2307/1267913},
  number = {1},
  journal = {Technometrics},
  publisher = {JSTOR},
  author = {Corbeil,  R. R. and Searle,  S. R.},
  year = {1976},
  month = feb,
  pages = {31}
}

@article{Srivastava2023,
  title = {Heritability Estimation Approaches Utilizing Genome‐Wide Data},
  volume = {3},
  ISSN = {2691-1299},
  url = {http://dx.doi.org/10.1002/cpz1.734},
  DOI = {10.1002/cpz1.734},
  number = {4},
  journal = {Current Protocols},
  publisher = {Wiley},
  author = {Srivastava,  Amit K. and Williams,  Scott M. and Zhang,  Ge},
  year = {2023},
  month = apr 
}

@article{zhou2012genome,
  title={Genome-wide efficient mixed-model analysis for association studies},
  author={Zhou, Xiang and Stephens, Matthew},
  journal={Nature genetics},
  volume={44},
  number={7},
  pages={821--824},
  year={2012},
  publisher={Nature Publishing Group},
  doi={10.1038/ng.2310}
}

@article{yang2011gcta,
  title={GCTA: a tool for genome-wide complex trait analysis},
  author={Yang, Jian and Lee, S Hong and Goddard, Michael E and Visscher, Peter M},
  journal={The American Journal of Human Genetics},
  year={2011},
  doi={10.1016/j.ajhg.2010.11.011}
}

@article{Speed2020,
  title = {Evaluating and improving heritability models using summary statistics},
  volume = {52},
  ISSN = {1546-1718},
  url = {http://dx.doi.org/10.1038/s41588-020-0600-y},
  DOI = {10.1038/s41588-020-0600-y},
  number = {4},
  journal = {Nature Genetics},
  publisher = {Springer Science and Business Media LLC},
  author = {Speed,  Doug and Holmes,  John and Balding,  David J.},
  year = {2020},
  month = mar,
  pages = {458–462}
}

@article{Zeng2017,
  title = {Non-parametric genetic prediction of complex traits with latent Dirichlet process regression models},
  volume = {8},
  ISSN = {2041-1723},
  url = {http://dx.doi.org/10.1038/s41467-017-00470-2},
  DOI = {10.1038/s41467-017-00470-2},
  number = {1},
  journal = {Nature Communications},
  publisher = {Springer Science and Business Media LLC},
  author = {Zeng,  Ping and Zhou,  Xiang},
  year = {2017},
  month = sep 
}

@article{bulik2015ldsc,
  title={LD Score regression distinguishes confounding from polygenicity in genome-wide association studies},
  author={Bulik-Sullivan, Brendan and Finucane, Hilary K and Anttila, Verneri and Gusev, Alexander and Day, Felix R and Loh, Po-Ru and ReproGen, Consortium and others},
  journal={Nature genetics},
  volume={47},
  number={3},
  pages={291--295},
  year={2015},
  publisher={Nature Publishing Group},
  doi={10.1038/ng.3211}
}

@article{Speed2018,
  title = {SumHer better estimates the SNP heritability of complex traits from summary statistics},
  volume = {51},
  ISSN = {1546-1718},
  url = {http://dx.doi.org/10.1038/s41588-018-0279-5},
  DOI = {10.1038/s41588-018-0279-5},
  number = {2},
  journal = {Nature Genetics},
  publisher = {Springer Science and Business Media LLC},
  author = {Speed,  Doug and Balding,  David J.},
  year = {2018},
  month = dec,
  pages = {277–284}
}

@article{Mayhew2017,
  title = {Assessing the Heritability of Complex Traits in Humans: Methodological Challenges and Opportunities},
  volume = {18},
  ISSN = {1389-2029},
  url = {http://dx.doi.org/10.2174/1389202918666170307161450},
  DOI = {10.2174/1389202918666170307161450},
  number = {4},
  journal = {Current Genomics},
  publisher = {Bentham Science Publishers Ltd.},
  author = {Mayhew,  Alexandra J. and Meyre,  David},
  year = {2017},
  month = jul 
}

@article{Zuk2012,
  title = {The mystery of missing heritability: Genetic interactions create phantom heritability},
  volume = {109},
  ISSN = {1091-6490},
  url = {http://dx.doi.org/10.1073/pnas.1119675109},
  DOI = {10.1073/pnas.1119675109},
  number = {4},
  journal = {Proceedings of the National Academy of Sciences},
  publisher = {Proceedings of the National Academy of Sciences},
  author = {Zuk,  Or and Hechter,  Eliana and Sunyaev,  Shamil R. and Lander,  Eric S.},
  year = {2012},
  month = jan,
  pages = {1193–1198}
}

@article{Heckerman2016,
  title = {Linear mixed model for heritability estimation that explicitly addresses environmental variation},
  volume = {113},
  ISSN = {1091-6490},
  url = {http://dx.doi.org/10.1073/pnas.1510497113},
  DOI = {10.1073/pnas.1510497113},
  number = {27},
  journal = {Proceedings of the National Academy of Sciences},
  publisher = {Proceedings of the National Academy of Sciences},
  author = {Heckerman,  David and Gurdasani,  Deepti and Kadie,  Carl and Pomilla,  Cristina and Carstensen,  Tommy and Martin,  Hilary and Ekoru,  Kenneth and Nsubuga,  Rebecca N. and Ssenyomo,  Gerald and Kamali,  Anatoli and Kaleebu,  Pontiano and Widmer,  Christian and Sandhu,  Manjinder S.},
  year = {2016},
  month = jul,
  pages = {7377–7382}
}

@article{Sun2019-lc,
  title     = "Heritability estimation and differential analysis of count data
               with generalized linear mixed models in genomic sequencing
               studies",
  author    = "Sun, Shiquan and Zhu, Jiaqiang and Mozaffari, Sahar and Ober,
               Carole and Chen, Mengjie and Zhou, Xiang",
  journal   = "Bioinformatics",
  publisher = "Oxford University Press (OUP)",
  volume    =  35,
  number    =  3,
  pages     = "487--496",
  month     =  feb,
  year      =  2019,
  copyright = "https://academic.oup.com/journals/pages/open\_access/funder\_policies/chorus/standard\_publication\_model",
  language  = "en"
}

@article{2024,
  volume = {53},
  ISSN = {1464-3685},
  url = {http://dx.doi.org/10.1093/ije/dyae039},
  DOI = {10.1093/ije/dyae039},
  number = {2},
  journal = {International Journal of Epidemiology},
  publisher = {Oxford University Press (OUP)},
  year = {2024},
  month = feb 
}

@article{Priv2020,
  title = {LDpred2: better,  faster,  stronger},
  volume = {36},
  ISSN = {1367-4811},
  url = {http://dx.doi.org/10.1093/bioinformatics/btaa1029},
  DOI = {10.1093/bioinformatics/btaa1029},
  number = {22–23},
  journal = {Bioinformatics},
  publisher = {Oxford University Press (OUP)},
  author = {Privé,  Florian and Arbel,  Julyan and Vilhjálmsson,  Bjarni J},
  editor = {Schwartz,  Russell},
  year = {2020},
  month = dec,
  pages = {5424–5431}
}

@inproceedings{Muneeb2022,
  series = {ICBBT 2022},
  title = {Heritability,  genetic variation,  and the number of risk SNPs effect on deep learning and polygenic risk scores AUC},
  url = {http://dx.doi.org/10.1145/3543377.3543387},
  DOI = {10.1145/3543377.3543387},
  booktitle = {2022 14th International Conference on Bioinformatics and Biomedical Technology},
  publisher = {ACM},
  author = {Muneeb,  Muhammad and Feng,  Samuel F. and Henschel,  Andreas},
  year = {2022},
  month = may,
  collection = {ICBBT 2022}
}

@article{Nustad2018-uy,
  title     = "A Bayesian mixed modeling approach for estimating heritability",
  author    = "Nustad, Haakon E and Page, Christian M and Reiner, Andrew H and
               Zucknick, Manuela and LeBlanc, Marissa",
  journal   = "BMC Proc.",
  publisher = "Springer Science and Business Media LLC",
  volume    =  12,
  number    = "Suppl 9",
  pages     = "31",
  month     =  sep,
  year      =  2018,
  language  = "en"
}

@article{Browning2011,
  title = {Population Structure Can Inflate SNP-Based Heritability Estimates},
  volume = {89},
  ISSN = {0002-9297},
  url = {http://dx.doi.org/10.1016/j.ajhg.2011.05.025},
  DOI = {10.1016/j.ajhg.2011.05.025},
  number = {1},
  journal = {The American Journal of Human Genetics},
  publisher = {Elsevier BV},
  author = {Browning,  Sharon R. and Browning,  Brian L.},
  year = {2011},
  month = jul,
  pages = {191–193}
}

@article{Manolio2009,
  title = {Finding the missing heritability of complex diseases},
  volume = {461},
  ISSN = {1476-4687},
  url = {http://dx.doi.org/10.1038/nature08494},
  DOI = {10.1038/nature08494},
  number = {7265},
  journal = {Nature},
  publisher = {Springer Science and Business Media LLC},
  author = {Manolio,  Teri A. and Collins,  Francis S. and Cox,  Nancy J. and Goldstein,  David B. and Hindorff,  Lucia A. and Hunter,  David J. and McCarthy,  Mark I. and Ramos,  Erin M. and Cardon,  Lon R. and Chakravarti,  Aravinda and Cho,  Judy H. and Guttmacher,  Alan E. and Kong,  Augustine and Kruglyak,  Leonid and Mardis,  Elaine and Rotimi,  Charles N. and Slatkin,  Montgomery and Valle,  David and Whittemore,  Alice S. and Boehnke,  Michael and Clark,  Andrew G. and Eichler,  Evan E. and Gibson,  Greg and Haines,  Jonathan L. and Mackay,  Trudy F. C. and McCarroll,  Steven A. and Visscher,  Peter M.},
  year = {2009},
  month = oct,
  pages = {747–753}
}

@article{Lee2011,
  title = {Estimating Missing Heritability for Disease from Genome-wide Association Studies},
  volume = {88},
  ISSN = {0002-9297},
  url = {http://dx.doi.org/10.1016/j.ajhg.2011.02.002},
  DOI = {10.1016/j.ajhg.2011.02.002},
  number = {3},
  journal = {The American Journal of Human Genetics},
  publisher = {Elsevier BV},
  author = {Lee,  Sang Hong and Wray,  Naomi R. and Goddard,  Michael E. and Visscher,  Peter M.},
  year = {2011},
  month = mar,
  pages = {294–305}
}

@article{ZHU20201557,
title = {Statistical methods for SNP heritability estimation and partition: A review},
journal = {Computational and Structural Biotechnology Journal},
volume = {18},
pages = {1557-1568},
year = {2020},
issn = {2001-0370},
doi = {https://doi.org/10.1016/j.csbj.2020.06.011},
url = {https://www.sciencedirect.com/science/article/pii/S2001037020303007},
author = {Huanhuan Zhu and Xiang Zhou}
}

@article{Collister2022,
  doi = {10.3389/fgene.2022.818574},
  url = {https://doi.org/10.3389/fgene.2022.818574},
  year = {2022},
  month = feb,
  publisher = {Frontiers Media {SA}},
  volume = {13},
  author = {Jennifer A. Collister and Xiaonan Liu and Lei Clifton},
  title = {Calculating Polygenic Risk Scores ({PRS}) in {UK} Biobank: A Practical Guide for Epidemiologists},
  journal = {Frontiers in Genetics}
}

@article{Lewis2020,
  doi = {10.1186/s13073-020-00742-5},
  url = {https://doi.org/10.1186/s13073-020-00742-5},
  year = {2020},
  month = may,
  publisher = {Springer Science and Business Media {LLC}},
  volume = {12},
  number = {1},
  author = {Cathryn M. Lewis and Evangelos Vassos},
  title = {Polygenic risk scores: from research tools to clinical instruments},
  journal = {Genome Medicine}
}

@article{Muneeb20221,
  doi = {10.21203/rs.3.rs-1298372/v1},
  url = {https://doi.org/10.21203/rs.3.rs-1298372/v1},
  year = {2022},
  month = feb,
  publisher = {Research Square Platform {LLC}},
  author = {Muhammad Muneeb and Samuel Feng and Andreas Henschel},
  title = {An empirical comparison between polygenic risk scores and machine learning for case/control classification}
}

@article{Yang2016,
  title = {GCTA-GREML accounts for linkage disequilibrium when estimating genetic variance from genome-wide SNPs},
  volume = {113},
  ISSN = {1091-6490},
  url = {http://dx.doi.org/10.1073/pnas.1602743113},
  DOI = {10.1073/pnas.1602743113},
  number = {32},
  journal = {Proceedings of the National Academy of Sciences},
  publisher = {Proceedings of the National Academy of Sciences},
  author = {Yang,  Jian and Lee,  S. Hong and Wray,  Naomi R. and Goddard,  Michael E. and Visscher,  Peter M.},
  year = {2016},
  month = jul 
}

@article{Choi2020,
  doi = {10.1038/s41596-020-0353-1},
  url = {https://doi.org/10.1038/s41596-020-0353-1},
  year = {2020},
  month = jul,
  publisher = {Springer Science and Business Media {LLC}},
  volume = {15},
  number = {9},
  pages = {2759--2772},
  author = {Shing Wan Choi and Timothy Shin-Heng Mak and Paul F. O'Reilly},
  title = {Tutorial: a guide to performing polygenic risk score analyses},
  journal = {Nature Protocols}
}

\end{document}